\documentclass[journal]{IEEEtran}
\usepackage{latexsym}
\usepackage{amsmath,epsfig}
\usepackage{enumitem}
\usepackage{multirow}
\usepackage{subfigure}
\usepackage{multicol}
\usepackage{booktabs}
\usepackage{bigstrut}
\usepackage{array}
\usepackage{amssymb}
\usepackage{url}
\usepackage{hyperref}
\usepackage{float}
\usepackage{indentfirst}
\usepackage{times}
\usepackage{psfrag}
\usepackage{cite}
\usepackage{textcomp}
\usepackage{multirow}
\usepackage{multicol}
\usepackage{stfloats}
\usepackage{xcolor}

\ifCLASSOPTIONcompsoc

  \usepackage[nocompress]{cite}
\else
  \usepackage{cite}
\fi

\ifCLASSINFOpdf

\else

\fi

\begin{document}

\title{Enhancing Quality for HEVC Compressed Videos}

\author{Ren Yang, \IEEEmembership{Student Member,~IEEE},
Mai Xu, \IEEEmembership{Senior Member,~IEEE}, Tie Liu, \\
Zulin Wang, \IEEEmembership{Member,~IEEE}, and Zhenyu Guan
\IEEEcompsocitemizethanks{\IEEEcompsocthanksitem Ren Yang, Mai Xu, Tie Liu, Zulin Wang and Zhenyu Guan are with Beihang University, China. This work is supported by NSFC under Grant number 61573037. Mai Xu is the corresponding author of this paper (e-mail: Maixu@buaa.edu.cn).}

}

\IEEEtitleabstractindextext{%

\begin{abstract}
The latest High Efficiency Video Coding (HEVC) standard has been increasingly applied to generate video streams over the Internet. However, HEVC compressed videos may incur severe quality degradation, particularly at low bit-rates. Thus, it is necessary to enhance the visual quality of HEVC videos at the decoder side. To this end, this paper proposes a Quality Enhancement Convolutional Neural Network (QE-CNN) method that does not require any modification of the encoder to achieve quality enhancement for HEVC. In particular, our QE-CNN method learns QE-CNN-I and QE-CNN-P models to reduce the distortion of HEVC I and P/B frames, respectively. The proposed method differs from the existing CNN-based quality enhancement approaches, which only handle intra-coding distortion and are thus not suitable for P/B frames. Our experimental results validate that our QE-CNN method is effective in enhancing quality for both I and P/B frames of HEVC videos. To apply our QE-CNN method in time-constrained scenarios, we further propose a Time-constrained Quality Enhancement Optimization (TQEO) scheme. Our TQEO scheme controls the computational time of QE-CNN to meet a target, meanwhile maximizing the quality enhancement. Next, the experimental results demonstrate the effectiveness of our TQEO scheme from the aspects of time control accuracy and quality enhancement under different time constraints. Finally, we design a prototype to implement our TQEO scheme in a real-time scenario.
\end{abstract}

}

\maketitle

\IEEEdisplaynontitleabstractindextext

\IEEEpeerreviewmaketitle

\ifCLASSOPTIONcompsoc
\IEEEraisesectionheading{\section{Introduction}\label{sec:introduction}}
\else
\vspace{-1.5em}
\section{Introduction}
\label{sec:introduction}
\fi

\IEEEPARstart{T}{he} High Efficiency Video Coding (HEVC) standard  \cite{sullivan2012overview} was officially issued in January 2013, and it has significantly improved the efficiency of video coding. HEVC is able to achieve a bit-rate saving of approximately $60\%$ with similar subjective quality \cite{Tan2016} compared with the previous H.264/AVC standard. However, when stored or transmitted at low bit-rates, HEVC videos also incur artifacts, such as blocking artifacts, ringing effects, blurring, and so forth. Such artifacts may cause severe degradation in the Quality of Experience (QoE) at the decoder side. Therefore, it is necessary to investigate how to enhance the visual quality of HEVC videos at the decoder side.

Over the past decade, there have been increasing interests in enhancing the visual quality of decoded images \cite{liew2004blocking,foi2007pointwise,jancsary2012loss,chang2014reducing,jung2012image,wang2013adaptive, dong2015compression,wang2016d3}. Recently, deep learning approaches have been successfully applied in enhancing the visual quality of decoded images. For example, Dong \textit{et al.} \cite{dong2015compression} designed a four-layer Convolutional Neural Network (CNN) \cite{lecun1998gradient}, named AR-CNN, for significantly improving the quality of JPEG images. Afterwards, other deep networks, e.g., $\text{\textbf{D}}^3$ \cite{wang2016d3} and TNRD \cite{Chen2017Trainable}, were proposed to reduce the JPEG artifacts.

For enhancing the quality of HEVC videos, some works \cite{han2015high,zhang2016structure,park2016cnn,dai2017convolutional} develop advanced postprocessing filters for the HEVC encoder. In particular, Park and Kim \cite{park2016cnn}, designed a CNN to replace the Sample Adaptive Offset (SAO) filter in the HEVC encoder and decoder. Later, based on AR-CNN \cite{dong2015compression}, Dai \textit{et al.} \cite{dai2017convolutional} proposed the VRCNN structure to replace the in-loop filters in HEVC intra-coding, achieving $4.6\%$ bit-rate reduction. However, \cite{dai2017convolutional} only handles the intra-mode coding of HEVC, in which the distortions of inter-coding frames, i.e., P/B frames, are not taken into consideration. More importantly, these approaches require modification at the encoder side, so that they are not suitable for enhancing the quality of already compressed videos.

Most recently, a deep network, called DCAD \cite{Wang2017A}, has been designed to improve the coding efficiency of HEVC by enhancing the quality of HEVC compressed videos at the decoder side. Nevertheless, the DCAD \cite{Wang2017A} approach uses the same CNN model to enhance both intra- and inter-coding frames, and the CNN model is learnt from HEVC compressed images only. Therefore, the DCAD \cite{Wang2017A} approach does not take the difference between intra- and inter-coding distortion into account, and thus it cannot effectively reduce the artifacts of both I and P/B frames. In this paper, we propose a novel Quality Enhancement CNN (QE-CNN) for enhancing HEVC quality at the decoder side. Our QE-CNN method has the specific architectures for HEVC intra- and inter-coding frames, respectively. Consequently, QE-CNN is able to extract distortion features of both intra- and inter-coding, and is effective for enhancing the quality of both I and P/B frames in HEVC videos. Moreover, it is worth pointing out that the HEVC encoder does not need to be modified when applying our QE-CNN method at the decoder side.

Unfortunately, the cost of visual quality enhancement is increased computational time \cite{dong2015compression,dai2017convolutional}, particularly for CNN-based approaches. Currently, the computational capacity of devices has dramatically improved, thus enabling video quality enhancement. However, the computational time is still limited for quality enhancement of HEVC, especially in real-time applications or for devices with low computational capacity. It is thus desirable to consider a time constraint when applying our QE-CNN method to enhance video quality for HEVC. To this end, we further propose a Time-constrained Quality Enhancement Optimization (TQEO) scheme. Specifically, we model the problem of TQEO by a formulation in which the quality enhancement of QE-CNN is maximized under the constraint of computational time. Then, two solutions are developed to solve our formulation for I and P/B frames, respectively. Consequently, the computational time of our QE-CNN method can be controlled to a target with optimal quality enhancement.

The novelty and contribution of our work are two-fold:

(1) The proposed QE-CNN method  is the first CNN-based method designed to learn the distortion features
of intra- and inter-coding, respectively. As such, the quality of both I and P/B frames can be effectively enhanced. Specifically, the convolutional layers in QE-CNN-I are designed to extract intra-coding features and enhance quality for intra-coding frames (i.e., I frames). Based on QE-CNN-I, QE-CNN-P is proposed to enhance the compression quality of inter-coding frames (i.e., P/B frames), by concatenating the intra- and inter-coding distortion features in a uniform network.

(2) Our TQEO scheme is a pioneering work to control the computational time of quality enhancement to a target. In our TQEO scheme, we first establish the TQEO formulation to maximize the MSE reduction of HEVC frames at the constraint on reduced computational complexity. Then, both compression and pixel domain features are adopted to predict the performance for quality enhancement of each CTU, and the computational complexity of QE-CNN-I/P is also modelled. Finally, we solve the TQEO formulation to decide whether to disable QE-CNN-I/P or replace QE-CNN-P by QE-CNN-I on each CTU. As such, the computational complexity can be controlled to a target with optimal quality enhancement.

This paper is an extended version of our conference paper \cite{Yang2017coding} presented in ICME 2017, with extensive
advanced works. Specifically, the training sets for QE-CNN-P is enlarged from 26 sequences in \cite{Yang2017coding} to 81 sequences in this paper. The QE-CNN model is advanced by adopting Parametric Rectified Linear Unit (PReLU) and residual learning, to achieve better quality performance than \cite{Yang2017coding} for both I and P/B frames. Most importantly, the TQEO scheme is proposed in this paper to control the computational time of our QE-CNN method in quality enhancement, and a prototype is designed to apply our TQEO scheme in a real-time quality enhancement scenario.

\vspace{-1em}
\section{Related works}

\subsection{Related works for quality enhancement}\label{sec:related1}
As mentioned above, over the past decade, many works  \cite{liew2004blocking,foi2007pointwise,wang2013adaptive,jancsary2012loss,chang2014reducing,jung2012image} have focused on enhancing the visual quality of images. Specifically, the method proposed by Liew \textit{et al.} \cite{liew2004blocking} reduces blocking artifacts of compressed images using an overcomplete wavelet representation. Foi \textit{et al.} \cite{foi2007pointwise} applied pointwise Shape-Adaptive Discrete Cosine Transform (SA-DCT) to reduce blocking and ringing effects caused by JPEG compression. Later, Wang \textit{et al.} \cite{wang2013adaptive} proposed filtering the boundaries between neighboring blocks for reducing blocky artifacts of JPEG images. Recently, Jancsary \textit{et al.} \cite{jancsary2012loss} achieved JPEG image deblocking by taking advantage of Regression Tree Fields (RTF). Moreover, some sparse coding methods exist for removing JPEG artifacts, such as \cite{chang2014reducing} and \cite{jung2012image}.

Over the past decade, deep learning has made impressive achievements in computer vision and image processing tasks \cite{krizhevsky2012imagenet,karpathy2014large,girshick2014rich,long2015fully}. Recently, deep learning has also been successfully applied to improve the visual quality of decoded images. Dong \textit{et al.} \cite{dong2015compression} proposed the four-layer AR-CNN to reduce the artifacts caused by JPEG coding. AR-CNN is designed based on a three-layer CNN used for super resolution, called SR-CNN \cite{Dong2014learning}. In \cite{Dong2014learning}, the functions of the three layers in SRCNN are defined as feature extraction, non-linear mapping and reconstruction. AR-CNN adds a feature denoising layer to SR-CNN to extract the features of JPEG distortion, thereby achieving quality enhancement on JPEG images. Then, some other deep networks, such as $\text{\textbf{D}}^3$ \cite{wang2016d3} and TNRD \cite{Chen2017Trainable}, are proposed to restore the JPEG distortion by utilizing the prior knowledge of JPEG compression. As reported in \cite{dong2015compression,wang2016d3}, these deep learning approaches significantly outperform other conventional methods, such as \cite{foi2007pointwise,jancsary2012loss,chang2014reducing}. The outstanding performance of \cite{dong2015compression,wang2016d3} highlights the promising application of deep learning in quality enhancement.

For videos, many works \cite{li2016lagrangian,gao2017temporal,han2015high,zhang2016structure,park2016cnn,dai2017convolutional} have been conducted to improve visual quality for the latest HEVC standard, aiming at improving the coding efficiency at the encoder side. The works of \cite{li2016lagrangian} and \cite{gao2017temporal} propose to improve the HEVC quality by Rate-Distortion Optimization (RDO) at the HEVC encoder. Han \textit{et al.} \cite{han2015high} developed a high-performance in-loop filter for HEVC, which is added after the original in-loop filter in the HEVC encoder and decoder.
Zhang \textit{et al.} \cite{zhang2016structure} proposed a Structure-driven Adaptive Non-local Filter (SANF), which is applied after the Deblocking Filter (DF) and before the SAO filter at both the HEVC encoder and decoder sides. Park and Kim \cite{park2016cnn} proposed replacing the SAO filter in HEVC by the re-trained SR-CNN \cite{Dong2014learning} to enhance the quality of HEVC videos. Later, based on AR-CNN \cite{dong2015compression}, Dai \textit{et al.} \cite{dai2017convolutional} proposed the VRCNN as an advanced in-loop filter in HEVC intra-coding. The VRCNN enhances the visual quality of HEVC without any increase in bit-rate compared with the conventional in-loop filters of HEVC. In other words, VRCNN is able to improve the coding efficiency of HEVC, i.e., a $4.6\%$ bit-rate reduction.
However, these approaches need to modify the HEVC encoder, thus not practical for existing HEVC bitstreams. Besides, VRCNN \cite{dai2017convolutional} is only designed for HEVC intra-coding mode, so that it is not suitable for inter-coding frames.

Most recently, Wang \textit{et al.} \cite{Wang2017A} have proposed a deep network to enhance the HEVC quality at the decoder side, which can be directly applied to the existing video streams. Nevertheless, in \cite{Wang2017A}, the same deep network structure, trained by HEVC intra-coded images, is used for enhancing the quality of both intra- and inter-coding frames. Therefore, the network of \cite{Wang2017A} is not able to deal with both intra- and inter-coding distortion, because there is no specific design for I and P/B frames, respectively. Actually, little attention has been devoted to reducing inter-coding distortion for further enhancing the visual quality of inter-frames.

\vspace{-1em}
\subsection{Related works on HEVC complexity control}\label{sec:related2}
Moreover, as far as we know, there are no works on controlling the computational time for video quality enhancement. The closest works are HEVC complexity control approaches, which aim at making the encoding or decoding time of HEVC meet a constraint. Specifically, at the encoder side, Corr{\^e}a \textit{et al.} \cite{correa2011complexity} proposed reducing the HEVC encoding complexity by skipping the computation on the Coding Tree Unit (CTU) partition at some frames. Instead, the CTU partitions of these frames are the same as those of previous frames. Then, the encoding complexity can be controlled by setting the number of frames that skip computation on the CTU partition. Later, Deng \textit{et al.} \cite{deng2014complexity,deng2016subjective} proposed a complexity control approach for HEVC encoding. In their approach, the maximal depths of some CTUs are limited to reduce the computational complexity to a target with minimal quality loss.

At the decoder side, Langroodi \emph{et al.} \cite{langroodi2015decoder} developed a complexity control approach for H.264/AVC decoding. In \cite{langroodi2015decoder}, the decoder sends its computational resource demand to the encoder side. Then, the settings of the encoder are adjusted according to the received demand such that the corresponding decoding complexity can be controlled. Rather than adjusting the encoder, Yang \emph{et al.} \cite{yang2016subjective,yang2016saliency} proposed an approach for controlling the HEVC decoding complexity. In their approach, the decoding complexity of HEVC is reduced by disabling DF and simplifying Motion Compensation (MC) at the expense of quality loss. Then, the HEVC decoding complexity is controlled to a certain target with minimal perceptual quality loss by disabling DF and simplifying the MC of non-salient CTUs. Similar to the above approaches, computational time control can also be studied for video quality enhancement under a time constraint. Unfortunately, to the best of our knowledge, no work in this direction exists.

\section{The proposed quality enhancement method}\label{QE}

In this section, we discuss the proposed QE-CNN method for enhancing the quality of decoded HEVC videos. In Section \ref{ARCNN}, we briefly review the overall architecture of AR-CNN, which is the basis of our QE-CNN method. In Sections \ref{CNNI} and \ref{CNNP}, we introduce the proposed network architecture of our QE-CNN method in detail, including QE-CNN-I for enhancing the visual quality of HEVC I frames and QE-CNN-P designed for HEVC P/B frames.

\vspace{-.5em}
\subsection{Overview of AR-CNN}\label{ARCNN}

In \cite{dong2015compression}, AR-CNN is developed to improve the visual quality of encoded JPEG images.  AR-CNN achieves a quality improvement on JPEG encoded images, and it can be viewed as the foundation of our method. Here, we briefly review the architecture of AR-CNN.

In AR-CNN, there are four convolutional layers without any pooling or fully connected layers. Specifically, as presented in \cite{dong2015compression}, the four layers of AR-CNN perform the functions of feature extraction, feature denoising, non-linear mapping and reconstruction. AR-CNN is designed as an end-to-end framework, which takes the JPEG compressed images as input and directly outputs the restored images. During training, all four layers of AR-CNN are jointly optimized.

The input image is denoted as $\text{\textbf{Y}}$, and the output of the $i$-th convolutional layer is defined as $F_i(\text{\textbf{Y}})$. Then, the architecture of AR-CNN can be expressed as
\begin{eqnarray}
\label{formulation1}
F_0(\text{\textbf{Y}})\!\!\!&=&\!\!\!\text{\textbf{Y}},\\
\label{formulation2}
F_{i}(\text{\textbf{Y}})\!\!\!&=&\!\!\!\max(0,W_i\ast F_{i-1}(\text{\textbf{Y}})+B_i),\ i\in\{1,2,3\},\\
\label{formulation3}
F_4(\text{\textbf{Y}})\!\!\!&=&\!\!\!W_4\ast F_3(\text{\textbf{Y}})+B_4,
\end{eqnarray}
where $W_i$ and $B_i$ are the weights and bias matrices of the $i$-th layer, respectively, and $\ast$ indicates the convolution operator. Note that $\max(0,x)$, known as Rectified Linear Unit (ReLU), is adopted in the first three layers as the non-linear activation function. The configuration of AR-CNN is summarized in Table \ref{tab:cfgarcnn}.

\begin{table}[!h]
\vspace{-1.5em}
\centering
\scriptsize
\caption{\footnotesize{Configuration of AR-CNN \cite{dong2015compression}}} \label{tab:cfgarcnn}
\vspace{-1em}
\begin{tabular}{|c|c|c|c|c|}
\hline
Layer index & Conv 1 & Conv 2 & Conv 3 & Conv 4 \\
\hline
Filter size & $9\times9$ & $7\times7$ & $1\times1$ & $5\times5$ \\
\hline
Filter number & 64 & 32 & 16 & 1 \\
\hline
$W$ learning rate & $10^{-4}$ & $10^{-4}$ & $10^{-4}$ &  $10^{-5}$ \\
\hline
$B$ learning rate & $10^{-5}$ & $10^{-5}$ & $10^{-5}$ &  $10^{-5}$ \\
\hline
\end{tabular}
\vspace{-1.5em}
\end{table}

The experimental results reported in \cite{dong2015compression} verify that AR-CNN has better performance than conventional methods, e.g., SA-DCT \cite{foi2007pointwise} and RTF \cite{jancsary2012loss}, for improving the quality of encoded images. Inspired by AR-CNN, we design QE-CNN for enhancing the quality of HEVC decoded videos, also taking advantage of the CNN.

\vspace{-.5em}
\subsection{The proposed QE-CNN-I}\label{CNNI}
We now focus on the proposed network of QE-CNN-I, which is able to handle distortion caused by intra-coding of HEVC, for enhancing the quality of HEVC I frames. Next, we present QE-CNN-I from the aspects of dataset, architecture and loss function.

\textbf{Dataset.} First, we establish a dataset for learning QE-CNN-I. In our work, the images of our training and validation sets for QE-CNN-I are the same as AR-CNN \cite{dong2015compression}. That is, the training and test sets (totally 400 images) of the BSDS500 database \cite{arbelaez2011contour} are used for training our QE-CNN-I model. The validation set (100 images) of BSDS500 is used for validation, non-overlapping with the training set.

Because QE-CNN-I aims to reduce the distortion of I frames in HEVC, we encode all training images using the HEVC all intra (AI) mode at QP = 32, 37, 42 and 47 using the default configuration in \textit{encoder\_intra\_main.cfg}. Before training, we decompose the ground-truth and HEVC encoded images into image patches with a size of $40 \times 40$, using a stride of 10. Consequently, the training set with 400 images provides a total of 522,000 pairs of training samples. Similarly, 34,500 pairs of validation samples are obtained.

Note that, the patch size is 40$\times$40 since the maximal TU size is 32$\times$32 and there is 8-pixel overlap for the 9$\times$9 convolutional filter. Here, we follow AR-CNN \cite{dong2015compression} to set the stride as 10. Since the stride is less than the patch size, a certain degree of redundancy exists. However, such overlapping generates a large number of training samples, which benefits the model training with sufficient data. In fact, choosing a stride is a trade-off between redundancy and the amount of training samples. According to our experiments, choosing the stride much larger than 10 reduces the performance of our method because of the lack on training samples. On the other hand, setting the stride less than 10 does not increase the performance according to our experiments, since too much redundancy is introduced.

\begin{table*}[!t]
\scriptsize
  \centering
  \vspace{-3.5em}
  \caption{Configuration and performance of AR-CNN and AR-CNN-1/2/3/4/5.}
  \vspace{-1em}
    \begin{tabular}{|c|c|c|c|c|c|c|c|c||c|c|}
    \hline
    QP   &   \multicolumn{8}{c||}{42} & 22\\
    \hline
    Network      & AR-CNN & AR-CNN-1 &  AR-CNN-2 & AR-CNN-3 & \multicolumn{2}{c|}{AR-CNN-4} & \multicolumn{2}{c||}{AR-CNN-5} & AR-CNN-3\\
    \hline
    Filter size & 9-7-1-5 & 9-7-1-5 &  \multicolumn{2}{c|}{9-7-3-1-5} & \multicolumn{2}{c|}{9-7-3-3-1-5}& \multicolumn{2}{c||}{9-7-3-3-3-1-5}& {9-7-3-1-5}\\
    \hline
    Filter number & 64-32-16-1 & 128-64-32-1  & \multicolumn{2}{c|}{128-64-64-32-1} & \multicolumn{2}{c|}{128-64-64-64-32-1}& \multicolumn{2}{c||}{128-64-64-64-64-32-1} & {128-64-64-32-1}\\
    \hline
    Layer number &  4 & 4 & \multicolumn{2}{c|}{5} & \multicolumn{2}{c|}{6} & \multicolumn{2}{c||}{7} & {5}\\
    \hline
    Function & ReLU & ReLU  & ReLU & PReLU & \ ReLU\ & PReLU & \ ReLU\ &PReLU & PReLU\\
    \hline
     $\Delta$PSNR (dB) & 0.2058 & 0.2203  & 0.2425 & \textbf{0.2487} & 0.2258 & 0.2435  & 0.2322 & 0.2348 & 0.2101\\
    \hline
    \end{tabular}%
    \vspace{-2em}
  \label{tab:filterno}%
\end{table*}%

\textbf{Architecture.} The architecture of QE-CNN-I is designed according to the following observations.

$\mathbf{\textit{\textbf{Observation\ 1.}}}$ The distortion caused by HEVC intra-coding is with more features than JPEG.

$\mathbf{\textit{\textbf{Analysis 1.}}}$ We analyze this observation from both theoretical and experimental aspects.  Theoretically, HEVC intra-coding is more complicated compared to JPEG. For example, HEVC supports different sizes of Discrete Cosine Transform (DCT), including $4\times4$, $8\times8$, $16\times16$ and $32\times32$  \cite{sullivan2012overview}, whereas JPEG only adopts $8\times8$ DCT \cite{wallace1992jpeg}. Moreover, the HEVC intra-picture prediction has 33 different directional orientations \cite{sullivan2012overview}, which is much more complex than the Direct-Current (DC) prediction in JPEG \cite{wallace1992jpeg}. Accordingly, more distortion sources
exist in HEVC intra-coding than in JPEG; thus, the distortion caused by HEVC intra-coding has more features than JPEG.

From the experimental aspect, we further analyze \textit{Observation 1} by testing CNN on the validation set under different configurations. First, based on AR-CNN, we design AR-CNN-1 with a larger number of convolutional filters. Then, we compare the performance of AR-CNN-1 with AR-CNN at QP = 42 on the validation set. The configuration of AR-CNN-1 and its performance are shown in Table \ref{tab:filterno}. As shown in Table \ref{tab:filterno}, AR-CNN-1, which has more filters, performs better than AR-CNN in terms of $\Delta$PSNR of HEVC I frames. Note that the results in Table \ref{tab:filterno} are evaluated on the validation set.

\begin{figure}[!t]
\centering
\hspace{-.4em}\subfigure[$W_1$ of AR-CNN trained on JPEG]{\includegraphics[width = .8\linewidth]{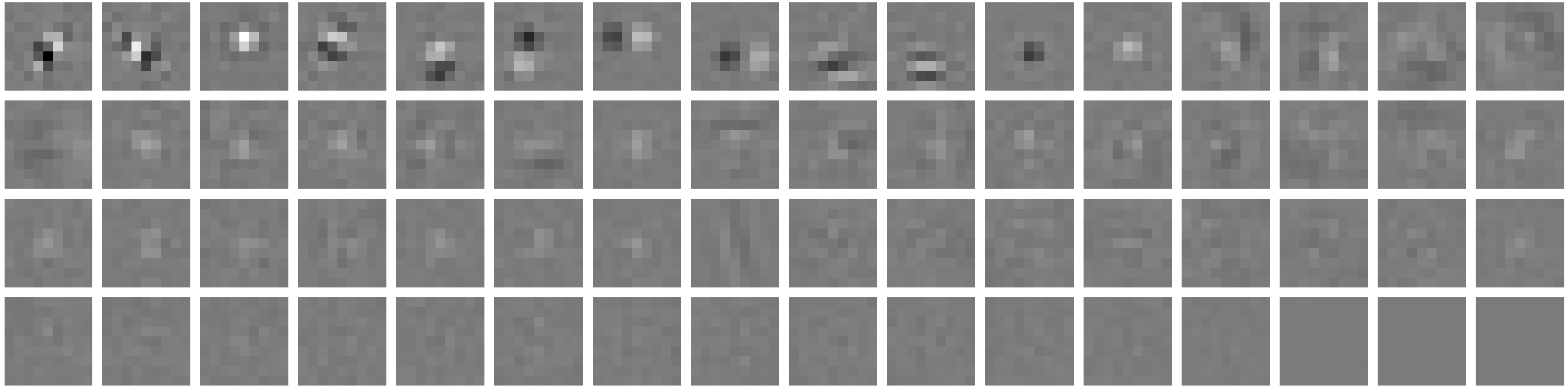}}\\
\vspace{-.5em}
\subfigure[$W_1$ of AR-CNN-1 trained on HEVC]{\includegraphics[width = .8\linewidth]{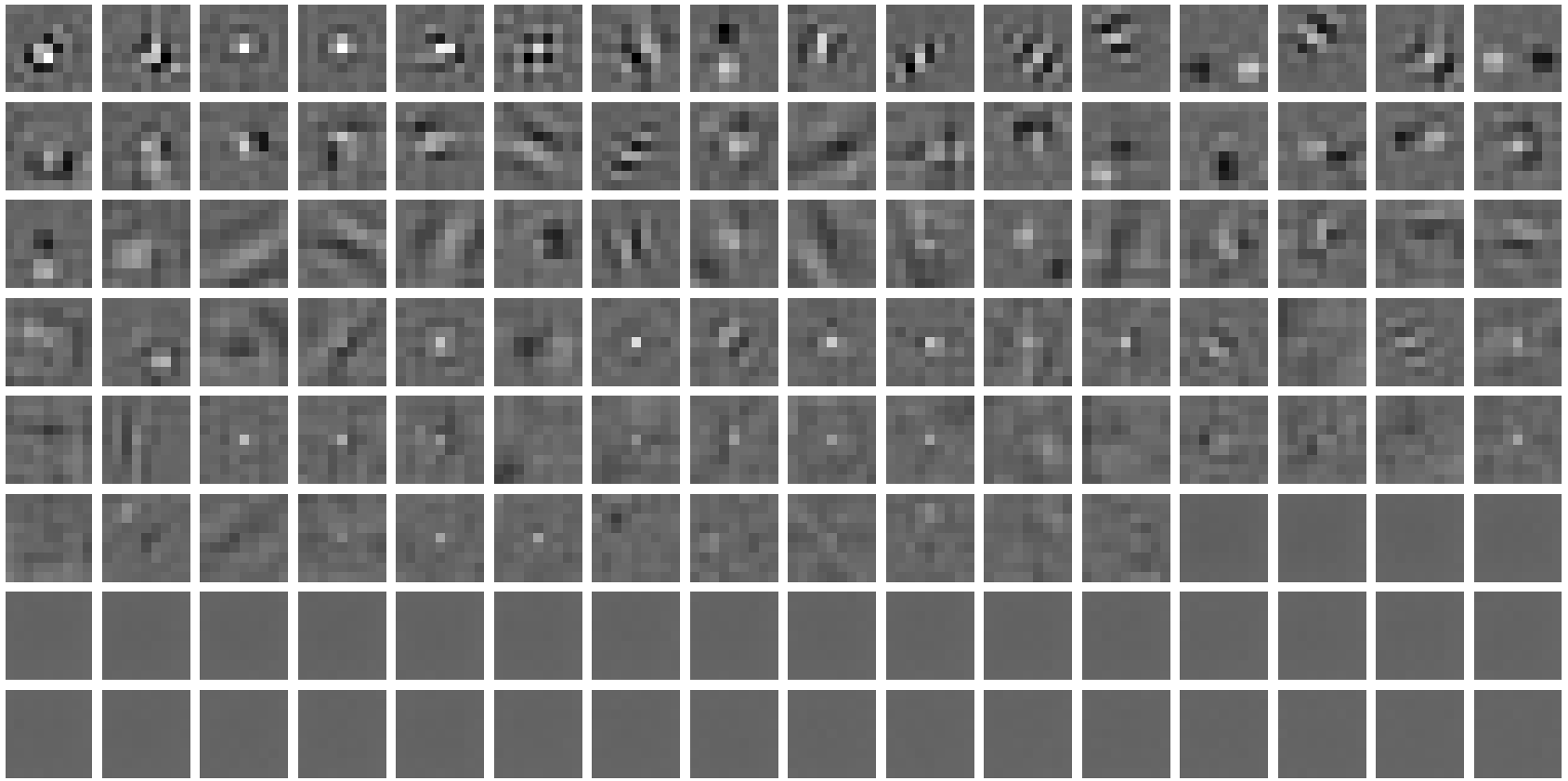}}
\vspace{-.5em}
\caption{Weight matrices of Conv 1 in AR-CNN and AR-CNN-1.}\label{fig:vis}
\vspace{-1em}
\end{figure}

Furthermore, Fig. \ref{fig:vis} presents the weight matrices of the first convolutional layer ($W_1$) in AR-CNN (trained by JPEG) and AR-CNN-1 (trained by HEVC) to visualize the distortion features of JPEG and HEVC. It can be seen from Fig. \ref{fig:vis}-(a) that the distortion features of JPEG are less than 64, because some of the 64 filters (in the last raw of Fig. \ref{fig:vis}-(a)) does not learn effective features. On the contrast, Fig. \ref{fig:vis}-(b) shows that HEVC videos are with much more distortion features than JPEG.
Finally, \textit{Observation 1} can be verified.
\hfill{$\blacksquare$}

Note that, the performance cannot be improved by further increasing the filter number. For example, setting filter numbers to be ``192-96-48-1''  and ``256-128-64-1'' results in $\Delta$PSNR = 0.2162 dB and 0.1873 dB, respectively. These results are worse than that of AR-CNN-1 (0.2203 dB). 

$\mathbf{\textit{\textbf{Observation\ 2.}}}$ AR-CNN-1 with one additional convolutional layer can extract more effective distortion-related features for HEVC, thus leading to better performance.

$\mathbf{\textit{\textbf{Analysis 2.}}}$ Recall that AR-CNN is a 4-layer network, in which Conv 1 is used to extract features and Conv 2 is for feature denoising \cite{dong2015compression}. For HEVC, we extend AR-CNN-1 to AR-CNN-2, which includes one additional layer after Conv 2 to further denoise the features. The configuration of AR-CNN-2 is shown in Table \ref{tab:filterno}. As such, AR-CNN-2 has 5 convolutional layers. Then, we test AR-CNN-2 on the validation set at QP = 42. As shown in Table \ref{tab:filterno}, AR-CNN-2 outperforms AR-CNN-1 for HEVC quality enhancement of I frames. Therefore, it can be confirmed that Conv 3 succeeds in further denoising the feature maps for HEVC encoded images.

However, Table \ref{tab:filterno} also shows that when more than one layer is added to AR-CNN-1, i.e., AR-CNN-4 and AR-CNN-5, the PSNR improvement decreases. This may be because too many additional layers introduce redundant parameters, leading to the over-fitting issue. To summarize, AR-CNN-1 with one additional convolutional layer achieves better performance. Finally, \textit{Observation 2} is validated. \hfill{$\blacksquare$}

$\mathbf{\textit{\textbf{Observation\ 3.}}}$ Parametric Rectified Linear Unit (PReLU), as the activation function instead of ReLU, can further improve the performance of AR-CNN-2.

$\mathbf{\textit{\textbf{Analysis 3.}}}$ PReLU is a learnable activation function, which is proved to have better performance than ReLU \cite{He2015Delving} and Leaky-ReLU. PReLU is defined as follows:
\begin{eqnarray}
\label{prelu}
\text{PReLU}(x) &=& \max(0,x)+a\cdot \min(0,x),
\end{eqnarray}
where $a$ is a parameter learned during the training stage. Recall that ReLU is defined as $\max(0,x)$; thus, its output always remains zero at the negative side. As a result, ReLU may incur the problem of ``dead features'' \cite{Zeiler2014Visualizing}. However, PReLU, whose slope is learnable under negative input, is able to avoid the problem of ``dead features''. We further test the performance of PReLU on the validation set at QP = 42. As shown in Table \ref{tab:filterno}, AR-CNN-3, which replaces ReLU by PReLU in AR-CNN-2, has better performance than AR-CNN-2. Besides, as Table II shows, AR-CNN-3 with PReLU also performs better than AR-CNN-2 using Leaky-ReLU. Consequently, \textit{Observation 3} can be validated. \hfill{$\blacksquare$}

\begin{figure}[!t]
  \centering
  \vspace{-2em}
  \includegraphics[width=1\linewidth]{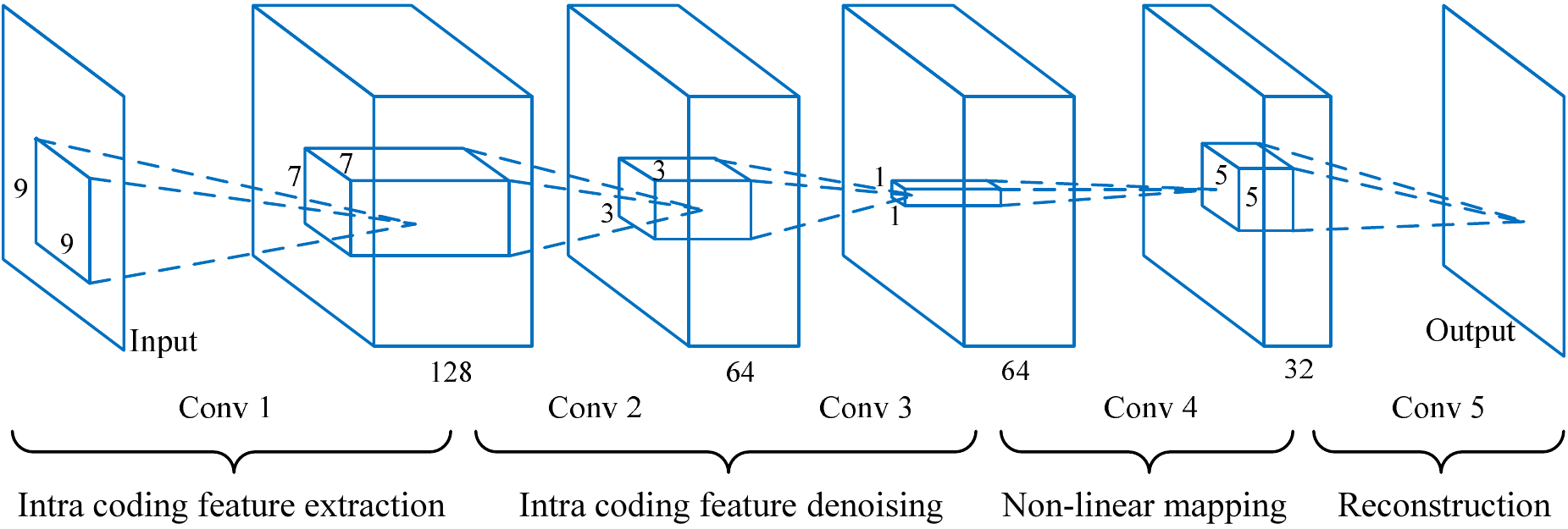}
  \vspace{-2em}
  \caption{\footnotesize{Network architecture of QE-CNN-I.}}\label{fig:arch}
\end{figure}

According to \textit{Observations 1}, \textit{2} and \textit{3}, we use AR-CNN-3 as our QE-CNN-I structure to enhance the quality of I frames of decoded HEVC videos. The architecture of QE-CNN-I is shown in Fig. \ref{fig:arch}, and its configuration is shown in Table \ref{tab:cfgescnn}. The formulation of QE-CNN-I is expressed as
\begin{eqnarray}
\label{ourf1}
F_0(\text{\textbf{Y}})\!\!\!\!\!&=&\!\!\!\!\!\text{\textbf{Y}},\\
\label{ourf2}
F_i(\text{\textbf{Y}})\!\!\!\!\!&=&\!\!\!\!\!\text{PReLU}(W_i\ast F_{i-1}(\text{\textbf{Y}})+B_i),\ i\in\{1,2,3,4\},\\
\label{ourf3}
F_5(\text{\textbf{Y}})\!\!\!\!\!&=&\!\!\!\!\!W_5\ast F_4(\text{\textbf{Y}})+B_5,
\end{eqnarray}
where $W_i$ and $B_i$ are the weights and bias matrices of the $i$-th layer, respectively. Note that PReLU is adopted as the non-linear activation function in the layers of Conv 1-4.

\textbf{Loss function.} We apply Mean Squared Error (MSE) as the loss function of our QE-CNN-I. Let $\{\text{\textbf{X}}_n\}_{n=1}^N$ be the set of raw image patches, seen as the ground-truth, and let $\{\text{\textbf{Y}}_n\}_{n=1}^N$ be patches of their corresponding compressed images. Here, $\{\text{\textbf{Y}}_n\}_{n=1}^N$ are input samples, whereas $\{\text{\textbf{X}}_n\}_{n=1}^N$ are the corresponding target output. Define $F(\cdot)$ as the output of QE-CNN-I. Then, the loss function is as follows:
\begin{eqnarray}
\label{loss}
L(\Theta) &=& \frac{1}{N} \sum_{n=1}^N{||F(\text{\textbf{Y}}_n; \Theta) - \text{\textbf{X}}_n ||_2^2},
\end{eqnarray}
where $\Theta = \{W_i, B_i\}$ stands for the weights and bias in QE-CNN-I. This loss function is minimized using the stochastic gradient descent algorithm with the standard Back-Propagation (BP). We follow \cite{dong2015compression} and \cite{dai2017convolutional} to set the batch size as 128 when training QE-CNN-I\footnote{Note that the batch size of 128 achieves the best result on both validation and test sets.}. Note that we first train our QE-CNN-I at QP = 42, and the networks for the other QPs (i.e., 22, 27, 32, 37 and 47) are fine-tuned from QP = 42.

\begin{table}[!t]
\centering
\vspace{-1em}
\scriptsize
\caption{\footnotesize{Configuration of QE-CNN-I.}} \label{tab:cfgescnn}
\vspace{-1em}
\begin{tabular}{|c|c|c|c|c|c|}
\hline
Layer & Conv 1& Conv 2& Conv 3& Conv 4& Conv 5\\
\hline
Filter size & $9\times9$ & $7\times7$ & $3\times3$ & $1\times1$ & $5\times5$ \\
\hline
Filter number & 128 & 64 & 64 & 32 & 1 \\
\hline
$W$ learning rate & $10^{-4}$ & $10^{-4}$ & $10^{-4}$ & $10^{-4}$ & $10^{-5}$ \\
\hline
$B$ learning rate & $10^{-5}$ & $10^{-5}$ & $10^{-5}$ & $10^{-5}$ & $10^{-5}$ \\
\hline
\end{tabular}
\vspace{-2em}
\end{table}

The validation result at QP = 22 is also shown in Table \ref{tab:filterno}. This result is comparable to that at QP = 42, indicating that our QE-CNN-I method is also effective for videos compressed at low QPs. Also, we evaluated the performance of applying the model trained at higher QP (QP = 27) to the sequences at QP = 22 from the validation set. However, only $\Delta$PSNR = 0.0049 dB can be achieved. This indicates that the model trained by low quality videos is not effective to high quality videos. Therefore, in this paper, we trained 6 QE-CNN models corresponding to QP = 22, 27, 32, 37, 42 and 47, respectively.

\subsection{The proposed QE-CNN-P}\label{CNNP}
Next, we introduce the proposed network of QE-CNN-P, which is designed to handle both HEVC intra- and inter-coding distortions. Therefore, it is able to enhance the quality of HEVC P/B frames. In the following, we present the dataset, architecture and loss function for QE-CNN-P.

\textbf{Dataset.} First, we establish a video database\footnote{Available at \url{https://github.com/ryangBUAA/Videos.git}.} that includes 89 sequences. We randomly selected 8 sequences (i.e., \textit{ParkRun}, \textit{Shields}, \textit{Stockholm}, \textit{Mobcal}, \textit{Tennis}, \textit{PartyScene}, \textit{FlowerVase} and \textit{BasketballDrill}) as our validation set for QE-CNN-P, and the other 81 sequences are used for training. Since our QE-CNN-P is implemented for enhancing the quality of P frames, all the training and validation sequences are encoded by HEVC Low-Delay P (LDP) mode at QP = 22, 27, 32, 37, 42 and 47, with the default configuration in \textit{encoder\_lowdelay\_P\_main.cfg}. We randomly select 10 P frames from each training sequence, and we decompose the raw and encoded frames into pairs of $40\times40$ image patches, with the stride being 15. In this way, we obtain 1,241,880 training sample pairs. Similarly, 279,090 validation sample pairs are obtained. Note that, we set larger stride for QE-CNN-P than QE-CNN-I to generate the training samples. It is because setting stride as 15 in QE-CNN-P already doubles the number of training samples of QE-CNN-I, and we do not need to set the stride to 10.

\begin{figure}[!t]
\vspace{-2.5em}
  \centering
  \includegraphics[width=1\linewidth]{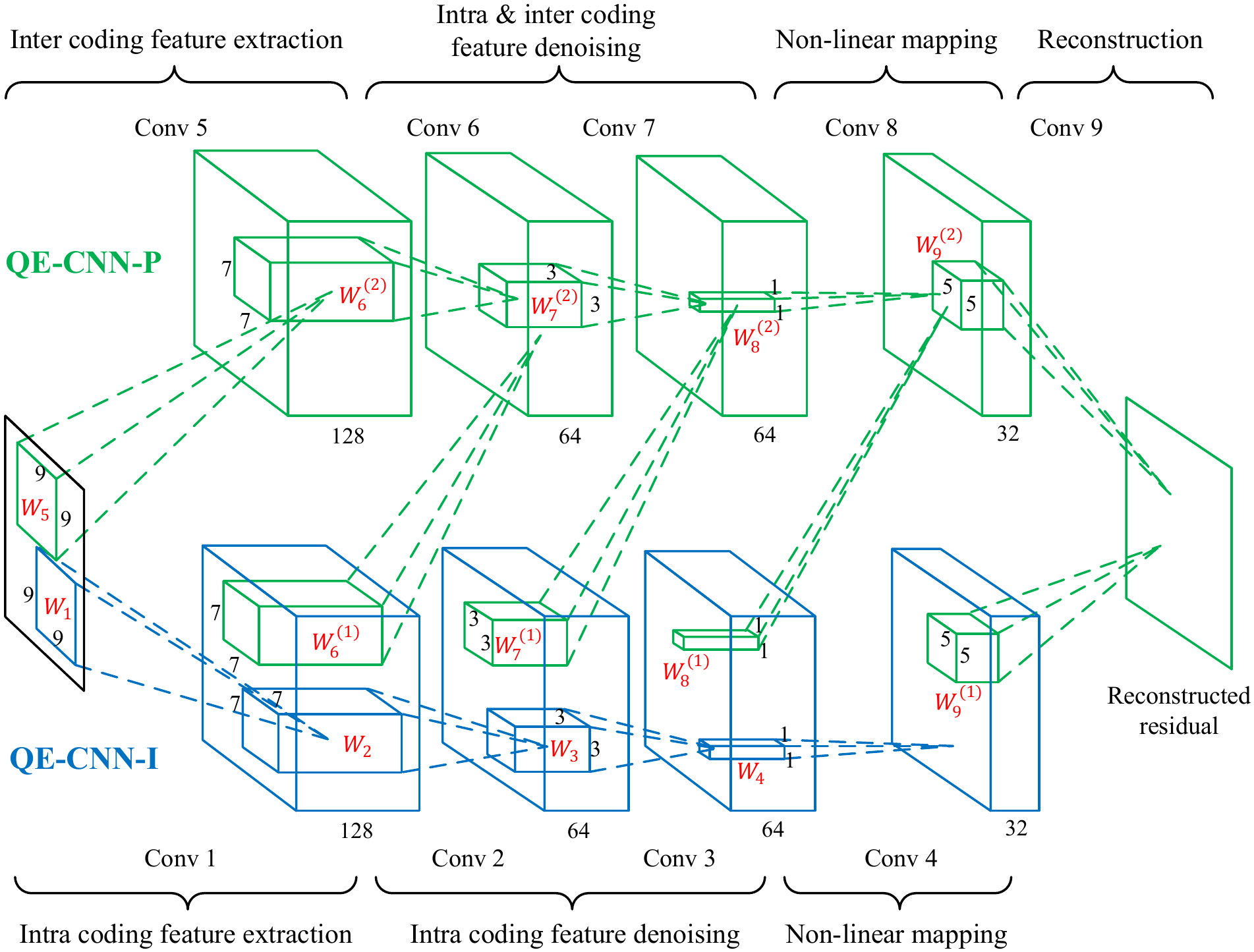}
  \vspace{-2em}
  \caption{\footnotesize{Network architecture of QE-CNN-P.}}\label{fig:arch2}
\end{figure}

\begin{table}[!t]
\centering
\vspace{-1em}
\scriptsize
\caption{\footnotesize{Configuration of QE-CNN-P.}} \label{tab:cfgescnn2}
\vspace{-1em}
\begin{tabular}{|c|c|c|c|c|c|}
\hline
\multirow{2}{*}{Layer} & Conv 1& Conv 2& Conv 3& Conv 4&  \\
       & Conv 5& Conv 6& Conv 7& Conv 8& Conv 9 \\
\hline
Filter size & $9\times9$ & $7\times7$ & $3\times3$ & $1\times1$ & $5\times5$ \\
\hline
Filter number & 128 & 64 & 64 & 32 & 1 \\
\hline
ILR for $W$  & $10^{-1}$ & $10^{-1}$ & $10^{-1}$ & $10^{-1}$ & $10^{-2}$ \\
\hline
ILR for $B$ & $10^{-2}$ & $10^{-2}$ & $10^{-2}$ & $10^{-2}$ & $10^{-2}$ \\
\hline
\end{tabular}
\vspace{-2em}
\end{table}

\textbf{Architecture.} In HEVC, only intra-picture coding is used in I frames, while both intra- and inter-coding are applied in P frames \cite{sullivan2012overview}. Thus, we learn the features of both HEVC intra- and inter-coding to enhance the visual quality of P frames. Accordingly, we design the QE-CNN-P containing 9 convolutional layers, as shown in Fig \ref{fig:arch2}. In QE-CNN-P, the architectures of layers Conv 1-4 are the same as those in QE-CNN-I, which are expected to extract and handle the distortion features of intra-coding. Moreover, Conv 5 is used to extract the distortion features of HEVC inter-coding. Recall that Conv 1 is used to extract intra-coding features. Then, the outputs of Conv 1 and Conv 5 are concatenated, and both are convolved by Conv 6. Thus, Conv 6 denoises the features of both intra- and inter-coding. Conv 7-9 in QE-CNN-P are designed in a similar way. Finally, Conv 9 in QE-CNN-P is used for reconstruction. Note that since Conv 1-3 in QE-CNN-P are used to extract intra-coding features, their parameters are fine-tuned from Conv 1-3 of the learned QE-CNN-I.

In Conv 6-9, $W^{(1)}_{i}$ and $W_i^{(2)}$ are defined as the weights of Conv $i$ ($i\in\{6,7,8,9\}$) used to convolve the feature maps of Conv 1-4 and Conv 5-8, respectively.  $\{W_i\}_{i=1}^5$ are also defined as the weights of Conv 1-5, and $\{B_i\}_{i=1}^9$ denotes the biases of Conv 1-9. Then, the formulation of QE-CNN-P can be expressed as
\begin{align}
&F_0(\text{\textbf{Y}})=\text{\textbf{Y}},\\
\label{ourf2}
&F_i(\text{\textbf{Y}})=\text{PReLU}(W_i\ast F_{i-1}(\text{\textbf{Y}})+B_i),\ i\in\{1,2,3,4\},\\
&F_5(\text{\textbf{Y}})=\text{PReLU}(W_5\ast F_0(\text{\textbf{Y}})+B_5),\\
&F_i(\text{\textbf{Y}})=\text{PReLU}(W^{(1)}_{i}\!\ast\! F_{i-5}(\text{\textbf{Y}})\! + \!W^{(2)}_{i}\!\ast\! F_{i-1}(\text{\textbf{Y}})\! +\!B_i),\nonumber\\
&\qquad \qquad \qquad \qquad \qquad \qquad \qquad \qquad \quad i\in\{6,7,8\},\\
&F_{9}(\text{\textbf{Y}})=W^{(1)}_{9}\!\ast\! F_{4}(\text{\textbf{Y}})\! + \!W^{(2)}_{9}\!\ast\! F_{8}(\text{\textbf{Y}})\! +\!B_9,
\end{align}
The sizes and numbers of filters at different layers in QE-CNN-P are shown in Table \ref{tab:cfgescnn2}, in which ILR means initial learning rate.

\textbf{Training procedure and loss function.} Since QE-CNN-P contains a larger number of layers than QE-CNN-I, we follow \cite{kim2016accurate,dai2017convolutional} to use the residual between raw and encoded videos as the ground-truth for training rather than the raw video itself. It has been verified in \cite{kim2016accurate} and \cite{dai2017convolutional} that learning the residual between raw and distorted frames achieves faster convergence and more accurate reconstruction. Moreover, we also apply learning rate decay in training QE-CNN-P. In this paper, the learning rate decays by a factor of 10 every 40 epochs. Meanwhile, to avoid exploding caused by a large learning rate $(\gamma)$, we also adopt the adjustable gradient clipping method \cite{kim2016accurate}. The gradient $g$ is clipped into the range of $g\leq|\beta/\gamma|$, where $\beta$ is a parameter for restricting the gradient. We follow \cite{kim2016accurate} to set the Initial Learning Rate (ILR) as $\gamma = 10^{-1}$ and set $\beta = 10^{-2}$.

For QE-CNN-P, we also apply MSE as the loss function. As residual learning is adapted in QE-CNN-P, its loss function, which is different from QE-CNN-I, is written as follows:
\begin{eqnarray}
\label{loss}
L(\Theta) &=& \frac{1}{N} \sum_{n=1}^N{||F(\text{\textbf{Y}}_n; \Theta) - (\text{\textbf{X}}_n-\text{\textbf{Y}}_n) ||_2^2},
\end{eqnarray}
where all notations are the same as in QE-CNN-I. The loss function is also minimized by stochastic gradient descent with the BP algorithm, and the batch size is set as 128. The same as QE-CNN-I, QE-CNN-P is first trained at QP = 42, and the other QPs are fine-tuned from QP = 42.

Note that due to the hierarchical coding structure, when compressing the sequences at a specific QP, the frame-level QPs in the sequences range from the sequence-level QP to sequence-level QP + 4. For example, the frame level QPs range from 32 to 36 when encoding at QP = 32, so that our QE-CNN model trained at QP = 32 is able to be applied to frames at QP = 32 to 36. As such, the 6 models trained at QP = 22, 27, 32, 37, 42 and 47 cover QPs from 22 to 51, and it is not necessary to train a model for each QP.

\section{Experimental results on quality enhancement}\label{EX-QE}
In this section, experimental results are presented to validate the effectiveness of our QE-CNN method in comparison with the latest quality enhancement methods\footnote{Here, we do not compare with $\text{\textbf{D}}^3$ \cite{wang2016d3} and TNRD \cite{Chen2017Trainable}. It is because $\text{\textbf{D}}^3$ and TNRD \cite{Chen2017Trainable}, which are designed according to prior knowledge of JPEG, cannot be implemented in HEVC.}: AR-CNN \cite{dong2015compression}, VRCNN \cite{dai2017convolutional} and DCAD \cite{Wang2017A}. In the following, we discuss the settings of our experiments in Section \ref{setting}, compare the performance of quality enhancement in Section \ref{quality}, and evaluate the computational time in Section \ref{ti}.

\begin{table*}[!t]
\vspace{-3.5em}
  \centering
  \scriptsize
  \caption{$\Delta$PSNR (dB) of our QE-CNN, AR-CNN \cite{dong2015compression}, VRCNN \cite{dai2017convolutional} and DCAD \cite{Wang2017A} methods.}
  \vspace{-1em}
\begin{tabular}{|p{.4cm}<{\centering}|p{.25cm}<{\centering}|p{.23cm}<{\centering}|c|c|c|c|c|c|c|c|}
\hline
\multirow{2}[2]{*}{QP} & \multirow{2}[2]{*}{\hspace{-.53em}Class} & \multirow{2}[2]{*}{\hspace{-.4em}Seq.} & \multicolumn{2}{c|}{AR-CNN \cite{dong2015compression}} & VRCNN \cite{dai2017convolutional}& \multicolumn{2}{c|}{DCAD \cite{Wang2017A}} & \multicolumn{2}{c|}{QE-CNN-I} & QE-CNN-P \\
\cline{4-11}      &       &       & I frames & P frames & I frames & I frames & P frames & I frames & P frames & P frames \\
\hline
\multirow{18}[4]{*}{42/32} & \multirow{2}[2]{*}{A} & 1     & 0.2372 / 0.2075 & 0.1499 / 0.1138 & 0.2643 / 0.2537 & 0.2311 / 0.2667 & 0.1390 / 0.1445 & \textbf{0.3423 / 0.3914} & 0.2504 / 0.2540 & \textbf{0.3285 / 0.3489} \\
\cline{3-11}      &       & 2     & 0.2896 / 0.1854 & 0.1330 / 0.1126 & 0.3294 / 0.1975 & 0.2877 / 0.2115 & 0.1271 / 0.1448 & \textbf{0.4461 / 0.3735} & 0.2501 / 0.2943 & \textbf{0.5649 / 0.5515} \\
\cline{2-11}      & \multirow{5}[2]{*}{B} & 3     & 0.3277 / 0.1103 & 0.2149 / 0.0704 & 0.3127 / 0.1325 & 0.3198 / 0.1462 & 0.2147 / 0.1065 & \textbf{0.4149 / 0.2199} & 0.3114 / 0.1689 & \textbf{0.3185 / 0.2160} \\
\cline{3-11}      &       & 4     & 0.1534 / 0.1021 & 0.1159 / 0.0755 & 0.1754 / 0.1357 & 0.2149 / 0.1510 & 0.1683 / 0.1163 & \textbf{0.2660 / 0.2259} & 0.2185 / 0.1798 & \textbf{0.2809 / 0.2287} \\
\cline{3-11}      &       & 5     & 0.1803 / 0.0541 & 0.1351 / 0.0611 & 0.1776 / 0.0862 & 0.2079 / 0.1083 & 0.1628 / 0.1161 & \textbf{0.2708 / 0.1714} & 0.2284 / 0.1818 & \textbf{0.3284 / 0.2698} \\
\cline{3-11}      &       & 6     & 0.1064 / 0.1421 & 0.0682 / 0.0742 & 0.1556 / 0.1525 & \textbf{0.1831} / 0.1481 & 0.1356 / 0.1104 & 0.1608 / \textbf{0.2442} & 0.1130 / 0.1435 & \textbf{0.1782 / 0.1826} \\
\cline{3-11}      &       & 7     & 0.1265 / 0.1086 & 0.1066 / 0.0915 & 0.1386 / 0.1777 & 0.1905 / 0.1334 & 0.1504 / 0.1196 & \textbf{0.1955 / 0.2212} & 0.1675 / 0.1917 & \textbf{0.3069 / 0.2917} \\
\cline{2-11}      & \multirow{4}[2]{*}{C} & 8     & 0.2711 / 0.1449 & 0.1781 / 0.0968 & 0.2946 / 0.1517 & 0.2722 / 0.2278 & 0.1770 / 0.1681 & \textbf{0.3624 / 0.3416} & 0.2636 / 0.2394 & \textbf{0.2878 / 0.2744} \\
\cline{3-11}      &       & 9     & 0.2371 / 0.1529 & 0.1572 / 0.1024 & 0.2663 / 0.1980 & 0.2716 / 0.2288 & 0.1761 / 0.1710 & \textbf{0.2891 / 0.2839} & 0.2114 / 0.2081 & \textbf{0.2790 / 0.2667} \\
\cline{3-11}      &       & 10    & 0.2046 / 0.1065 & 0.1614 / 0.0937 & 0.2351 / 0.2196 & 0.2480 / 0.2043 & 0.1801 / 0.1682 & \textbf{0.2925 / 0.2411} & 0.2571 / 0.2111 & \textbf{0.2659 / 0.2554} \\
\cline{3-11}      &       & 11    & 0.1826 / 0.1157 & 0.1843 / 0.1079 & 0.2616 / 0.2092 & 0.0510 / 0.1802 & 0.0773 / 0.1646 & \textbf{0.2698 / 0.2748} & 0.2386 / 0.2401 & \textbf{0.2837 / 0.3079} \\
\cline{2-11}      & \multirow{3}[2]{*}{E} & 12    & 0.2876 / 0.1367 & 0.1664 / 0.0899 & 0.2823 / 0.2077 & 0.2620 / 0.2145 & 0.1389 / 0.1424 & \textbf{0.3807 / 0.3767} & 0.2761 / 0.2795 & \textbf{0.2994 / 0.3411} \\
\cline{3-11}      &       & 13    & 0.3812 / 0.2592 & 0.2955 / 0.1809 & 0.4060 / 0.3640 & 0.3777 / 0.3678 & 0.2929 / 0.2526 & \textbf{0.4949 / 0.5486} & 0.4137 / 0.3884 & \textbf{0.4636 / 0.4817} \\
\cline{3-11}      &       & 14    & 0.4004 / 0.2146 & 0.2943 / 0.1677 & 0.4303 / 0.2866 & 0.4124 / 0.3073 & 0.3137 / 0.2395 & \textbf{0.5065 / 0.4732} & 0.4020 / 0.3601 & \textbf{0.4380 / 0.4324} \\
\cline{2-11}      & \multirow{3}[2]{*}{E'} & 15    & 0.3419 / 0.2222 & 0.2289 / 0.1497 & 0.3619 / 0.2897 & 0.3420 / 0.3094 & 0.2225 / 0.2070 & \textbf{0.4585 / 0.4615} & 0.3466 / 0.3295 & \textbf{0.4147 / 0.4545} \\
\cline{3-11}      &       & 16    & 0.2744 / 0.2114 & 0.1415 / 0.1123 & 0.2126 / 0.2644 & 0.2606 / 0.2855 & 0.1317 / 0.1438 & \textbf{0.3802 / 0.4576} & 0.2769 / 0.3200 & \textbf{0.4181 / 0.5053} \\
\cline{3-11}      &       & 17    & 0.2528 / 0.1644 & 0.1689 / 0.1111 & 0.2730 / 0.2215 & 0.1363 / 0.2392 & 0.0414 / 0.1629 & \textbf{0.3655 / 0.3418} & 0.2802 / 0.2541 & \textbf{0.3361 / 0.3300} \\
\cline{2-11}      & \multicolumn{2}{c|}{\textbf{Average}} & 0.2503 / 0.1552 & 0.1706 / 0.1066 & 0.2693 / 0.2087 & 0.2511 / 0.2194 & 0.1676 / 0.1576 & \textbf{0.3469 / 0.3323} & 0.2650 / 0.2496 & \textbf{0.3407 / 0.3367} \\
\hline
\hline
22/27& \multicolumn{2}{c|}{\textbf{Average}} & 0.1194 / 0.1750 & 0.1416 / 0.1545 &  0.1552 / 0.2112 & 0.2119 / 0.3071 & 0.2513 / 0.2891 & \textbf{0.2622 / 0.3399} & 0.2820  / 0.3055   & \textbf{0.3241 / 0.3416} \\
\hline
\multicolumn{11}{c}{1: \textit{Traffic} 2:\textit{PeopleOnStreet} 3: \textit{BQTerrace} 4: \textit{Cactus} 5: \textit{BasketballDrive} 6: \textit{ParkScene} 7: \textit{Kimono} } \\

\multicolumn{11}{c}{8: \textit{BQMall} 9: \textit{RaceHorses} 10: \textit{Keiba} 11: \textit{Mobisode} 12: \textit{Johnny}  13: \textit{FourPeople} 14: \textit{KristenAndSara} 15: \textit{Vidyo1} 16: \textit{Vidyo3} 17: \textit{Vidyo4} }
    \end{tabular}%
  \label{tab:results1}%
  \vspace{-2em}
\end{table*}%

\vspace{-.8em}
\subsection{Settings}\label{setting}

In our experiments, we test our method and the other three methods on 17 sequences from the JCT-VC \cite{bossen2011common} database. Details about the test sequences are presented in Table \ref{tab:results1}. The training and validation sequences are the same as those introduced in Sections \ref{CNNI} and \ref{CNNP}. Note that the test sequences do not overlap with those of the training and validation sets. For each sequence, the corresponding HEVC bitstream is generated by HM 16.0 with LDP mode at six QP values (i.e., 22, 27, 32, 37, 42 and 47). Here, the default configuration file \textit{encoder\_lowdelay\_P\_main.cfg} is used. Then, our QE-CNN method and the AR-CNN, VRCNN and DCAD methods are applied to enhance the quality of the encoded HEVC sequences. Besides, we further transfer our QE-CNN method to HEVC Random Access (RA) sequences to verify the generalization capability.

Unlike the training, the frames are not divided into small patches in the test. Instead, the whole frame is fed into QE-CNN. All layers in our QE-CNN structure are convolutional layers, and thus the input size of test images are not the same as that of training patches (i.e., $40\times40$). It is because for convolutional layers, only the weights and bias of convolutional kernels (filters) are learnt during training, and the learnt kernels can be used to convolve input frames with any size. As a result, the output is of the same size as the input.

Note that for a fair comparison, AR-CNN, which is used for JPEG deblocking, is re-trained on HEVC compressed training samples. Recall that the hyperparameters of our QE-CNN method, e.g., learning rate, batch size, and so forth, are introduced in Section \ref{QE}.

\vspace{-.8em}
\subsection{Performance of quality enhancement}\label{quality}

We now evaluate the performance of our QE-CNN method in terms of quality enhancement, comparing with the conventional AR-CNN \cite{dong2015compression}, VRCNN \cite{dai2017convolutional} and DCAD \cite{Wang2017A}. The quality enhancement is measured by Y-PSNR improvement ($\Delta$PSNR), and the $\Delta$PSNR results of our method and the other three methods are reported in Table \ref{tab:results1}.

\textbf{Quality enhancement on I frames.} As shown in Table \ref{tab:results1}, the proposed QE-CNN-I model clearly outperforms AR-CNN, VRCNN and DCAD on I frames over all test sequences. At QP = 42, the average $\Delta$PSNR (0.3469 dB) of our QE-CNN-I model obviously outperforms AR-CNN (0.2503 dB), VRCNN (0.2693 dB) and DCAD (0.2511 dB).
In particular, QE-CNN-I has up to a 0.5065 dB Y-PSNR improvement on I frames, which is achieved in sequence \textit{KristenAndSara}. At QP = 32, our QE-CNN-I model  enhances the quality of I frames by 0.3328 dB on average, doubling the enhancement of AR-CNN (0.1552 dB). Meanwhile, the quality enhancement of QE-CNN-I is $59.44\%$ higher than that of VRCNN (0.2087 dB) and $51.69\%$ higher than that of DCAD (0.2194 dB). The maximum enhancement of QE-CNN-I is 0.5486 dB, which is achieved in sequence \textit{FourPeople}. Comparable results can be found for QP = 22 and 27 in Table \ref{tab:results1}, indicating that our QE-CNN-I model is still effective for lower QPs outperforming the conventional methods. To summarize, the proposed QE-CNN-I model significantly outperforms the state-of-the-art AR-CNN, VRCNN and DCAD methods in enhancing the quality of I frames.

\begin{table}[!t]
\vspace{-1em}
  \centering
  \small
  \caption{BD-rate $(\%)$ results of over HM 16.0.}
  \vspace{-1em}
   \begin{tabular}{|c|c|c|c|c|}
    \multicolumn{5}{c}{\footnotesize{(a) BD-rate $(\%)$ calculated at QP = 22, 27, 32 and 37.}}\\
\hline
Class &
  Seq. &
  AR-CNN \cite{dong2015compression} &
  DCAD \cite{Wang2017A} &
  QE-CNN
  \\
\hline
\multirow{2}[2]{*}{A} &
  1 &
  -5.2822  &
  -7.8811  &
  \textbf{-10.9653 }
  \\
\cline{2-5} &
  2 &
  -4.7737  &
  -6.3344  &
  \textbf{-11.2186 }
  \\
\hline
\multirow{5}[2]{*}{B} &
  3 &
  -3.7115  &
  -4.2143  &
  \textbf{-9.7245 }
  \\
\cline{2-5} &
  4 &
  -4.1001  &
  -6.2853  &
  \textbf{-10.1089 }
  \\
\cline{2-5} &
  5 &
  -2.8679  &
  -3.8049  &
  \textbf{-8.1896 }
  \\
\cline{2-5} &
  6 &
  -4.2460  &
  -6.0328  &
  \textbf{-9.2608 }
  \\
\cline{2-5} &
  7 &
  -5.4054  &
  -8.4259  &
  \textbf{-10.9834 }
  \\
\hline
\multirow{4}[2]{*}{C} &
  8 &
  -3.5527  &
  -5.2386  &
  \textbf{-6.7784 }
  \\
\cline{2-5} &
  9 &
  -3.8050  &
  -4.8589  &
  \textbf{-7.3078 }
  \\
\cline{2-5} &
  10 &
  -3.4831  &
  -5.3648  &
  \textbf{-7.0244 }
  \\
\cline{2-5} &
  11 &
  -5.0319  &
  -7.9474  &
  \textbf{-11.8583 }
  \\
\hline
\multirow{3}[2]{*}{E} &
  12 &
  -6.0658  &
  -13.7325  &
  \textbf{-14.4719 }
  \\
\cline{2-5} &
  13 &
  -6.4344  &
  -10.3018  &
  \textbf{-13.6474 }
  \\
\cline{2-5} &
  14 &
  -6.6520  &
  -9.5921  &
  \textbf{-12.7502 }
  \\
\hline
\multirow{3}[2]{*}{E'} &
  15 &
  -6.6232  &
  -11.6557  &
  \textbf{-14.8835 }
  \\
\cline{2-5} &
  16 &
  -7.3530  &
  -13.9071  &
  \textbf{-16.3591 }
  \\
\cline{2-5} &
  17 &
  -4.9210  &
  -10.2085  &
  \textbf{-12.4934 }
  \\
\hline
\multicolumn{2}{|c|}{\textbf{Average}} &
  -4.9593  &
  -7.9874  &
  \textbf{-11.0603 }
  \\
\hline

%


    \multicolumn{5}{c}{\footnotesize{(b) BD-rate $(\%)$ calculated at QP = 32, 37, 42 and 47.}}\\
   \hline
    \multicolumn{2}{|c|}{\textbf{AVERAGE}} &
      -3.9552 & -4.2948 &
      \textbf{-8.3085}
      \\
    \hline
\multicolumn{5}{c}{\scriptsize{1: \textit{Traffic} 2: \textit{PeopleOnStreet} 3: \textit{BQTerrace} 4: \textit{Cactus} 5: \textit{BasketballDrive}}}\\

\multicolumn{5}{c}{\scriptsize{6: \textit{ParkScene} 7: \textit{Kimono} 8: \textit{BQMall} 9: \textit{RaceHorses} 10: \textit{Keiba} 11: \textit{Mobisode}}}\\
\multicolumn{5}{c}{\scriptsize{12: \textit{Johnny}  13: \textit{FourPeople} 14: \textit{KristenAndSara} 15: \textit{Vidyo1} 16: \textit{Vidyo3} 17: \textit{Vidyo4}}}\\

    \end{tabular}%
  \label{tab:BD}%
  \vspace{-2em}
\end{table}%

\textbf{Quality enhancement on P frames.} As shown in Table \ref{tab:results1}, the QE-CNN-I and QE-CNN-P models both perform better than AR-CNN and DCAD over P frames of all test sequences. Note that we do not compare with VRCNN over HEVC P frames because VRCNN \cite{dai2017convolutional} is only designed to improve the coding efficiency of HEVC intra-mode, and applying VRCNN on P frames requires modification of the HEVC encoder. Table \ref{tab:results1} shows that QE-CNN-I also has the ability to enhance the quality of P frames due to the reduction in intra-coding distortion. At QP = 42, QE-CNN-I still achieves a $55.36\%$ and $58.11\%$ gain of $\Delta$PSNR (0.2650 dB) over AR-CNN (0.1706 dB) and DCAD (0.1676 dB), respectively. Similarly, at QP = 32, the $\Delta$PSNR of QE-CNN-I is 0.2496 dB for P frames, which is much higher than 0.1066 dB of AR-CNN and 0.1576 dB of DCAD.

Furthermore, our QE-CNN-P model yields better quality than QE-CNN-I for the P frames of HEVC sequences, benefiting from its specific design in enhancing the quality of P frames. At QP = 42, QE-CNN-P averagely reaches a 0.3407 dB Y-PSNR improvement on P frames, which is $28.57\%$ higher than that of QE-CNN-I (0.2650 dB). Moreover, our QE-CNN-P model doubles the PSNR improvement of AR-CNN (0.1706 dB) and DCAD (0.1676). At QP = 32, our QE-CNN-P model improves Y-PSNR by 0.3328 dB on average, which is also considerably higher than that of QE-CNN-I (0.2496 dB). More importantly, the QE-CNN-P model triples the quality enhancement of AR-CNN (0.1066 dB) and doubles that of DCAD (0.1576 dB) for P frames. Similar results can be found at QP = 37 and 47. In other words, for the P frames of HEVC sequences, our QE-CNN-P model improves the quality enhancement performance of both compared methods (i.e., AR-CNN and DCAD) and our QE-CNN-I model. At QP = 22 and 27, similar results can be observed from Table \ref{tab:results1}.

Note that, our QE-CNN-P model also performs well at scene change frames, at which most of coding units are encoded by intra mode. For example, The scene change frame of \textit{Kimono} (i.e., the 141-st frame) has $\Delta$PSNR = 0.2238 dB and 0.2216 dB at QP = 32 and 37, respectively. These results are comparable to the average quality improvement for all P frames after scene change, i.e., 0.2278 dB at QP = 32 and 0.2362 dB at QP = 37. Besides, the QE-CNN-P model also achieves good performance for I frames, which is comparable to that for P frmes. For example, for all test sequences compressed at QP = 37, QE-CNN-P has $\Delta$PSNR = 0.3421 dB at I frames and $\Delta$PSNR = 0.3586 dB at P frames  on average. This further verifies that the QE-CNN-P model has the ability to handle both intra- and inter-coding distortion.

%

\textbf{Subjective quality performance.} 
Additionally, we conduct a subjective experiment to evaluate the subjective quality of our QE-CNN method and the conventional methods of AR-CNN \cite{dong2015compression} and DCAD \cite{Wang2017A}. In our experiments, the Difference Mean Opinion Score (DMOS) test was conducted to rate subjective quality difference between raw and enhanced HEVC frames, by the means of Single Stimulus Continuous Quality Score (SSCQS), adopted Rec. ITU-R BT.500 \cite{recommendation2002500}. There are totally 12 non-expert subjects involved in the test. During the test, sequences were displayed at random order. After viewing each sequence, the subjects were asked to rate the subjective score. The rating score includes excellent (100-81), good (80-61), fair (60-41), poor (40-21), and bad (20-1). As a result, DMOS value of each enhanced sequence can be calculated to measure the difference of subjective quality between the raw sequences and sequences enhanced by our QE-CNN method or AR-CNN \cite{dong2015compression} and DCAD \cite{Wang2017A}. It worth pointing out that the smaller DMOS value represents the better subjective quality.

Table \ref{tab:sub} shows the DMOS scores of three methods for all test sequences at QP = 37, in which ``QE-CNN (50\%)'' indicates our QE-CNN with 50\% computational complexity after applying our TQEO scheme. As Table \ref{tab:sub} shows, our QE-CNN achieves better subjective quality than \cite{dong2015compression} and \cite{Wang2017A} on 15 out of 17 test sequences. The averaged DMOS score of our QE-CNN is 44.49, much smaller than that of AR-CNN (54.42) and DCAD (51.86). This validates the better subjective quality achieved by our QE-CNN method compared with the existing methods AR-CNN \cite{dong2015compression} and DCAD \cite{Wang2017A}. Besides, some selected frames enhanced by AR-CNN \cite{dong2015compression}, DCAD \cite{Wang2017A} and our QE-CNN methods are shown in the \textit{Supporting Document} to subjectively present the performance of quality enhancement.

\begin{table}[!t]
  \centering
  \scriptsize
  \vspace{-4em}
  \caption{DMOS/sharpness difference scores at QP = 37.}
  \vspace{-1.5em}
     \begin{tabular}{|c|c|c|c|c||c|}
    \hline
     &   & AR-CNN \cite{dong2015compression} & DCAD \cite{Wang2017A} & QE-CNN & QE-CNN (50\%) \\
    \hline
    \multirow{2}[2]{*}{A} & 1     & 43.72 / 0.25  & 41.59 / 0.25 & \textbf{35.72} / 0.25 & 45.08 / 0.25  \\
\cline{2-6}          & 2     & 45.95 / 0.00  & 35.17 / 0.00  & \textbf{26.73} / 0.00 & 34.82 / 0.00 \\
    \hline
    \multirow{5}[2]{*}{B} & 3     & 39.55 / 0.00  & 39.78 / 0.00  & \textbf{31.71} / 0.00 & 38.15 / 0.00 \\
\cline{2-6}          & 4     & 52.88 / 0.83  & \textbf{47.38} / 0.92 & 48.64 / 1.00  & 48.02 / 0.83 \\
\cline{2-6}          & 5     & 54.79 / 0.17 & 49.67 / 0.17 & \textbf{38.34} / 0.17 & 45.83 / 0.17 \\
\cline{2-6}          & 6     & 58.21 / 0.42  & 55.74 / 0.25 & \textbf{48.09} / 0.33 & 65.23 / 0.33 \\
\cline{2-6}          & 7     & 56.35 / 0.50 & 52.40 / 0.42 & \textbf{42.59} / 0.50 & 50.42 / 0.33\\
    \hline
    \multirow{4}[2]{*}{C} & 8     & 59.30 / 0.42 & 50.82 / 0.50 & \textbf{46.95} / 0.50 & 48.46 / 0.33 \\
\cline{2-6}          & 9     & 49.07 / 1.00& 54.52 / 1.00  & \textbf{42.87} / 1.00 & 52.19 / 1.00  \\
\cline{2-6}          & 10    & 45.71 / 0.08  & 37.61 / 0.08  & 34.15 / 0.00 & \textbf{33.47} / 0.00 \\
\cline{2-6}          & 11    & 57.53 / 0.33  & 59.68 / 0.33  & \textbf{53.58} / 0.17 & 55.93 / 0.25 \\
    \hline
    \multirow{3}[2]{*}{E} & 12    & 61.02 / 0.42  & 50.74 / 0.50 & \textbf{48.57} / 0.17 & 53.56 / 0.33  \\
\cline{2-6}          & 13    & 61.89 / 0.58  & 61.39 / 0.50 & \textbf{55.31} / 0.58 & 61.84 / 0.58  \\
\cline{2-6}          & 14    & 61.57 / 0.25  & 60.39 / 0.42  & \textbf{45.84} / 0.33 & 55.02 / 0.33 \\
    \hline
    \multirow{3}[2]{*}{E'} & 15    & 57.63 / 0.42  & 60.99 / 0.50  & 48.04 / 0.50 & \textbf{43.46} / 0.42\\
\cline{2-6}          & 16    & 63.67 / 0.83 & 63.41 / 0.83 & 63.87  / 0.67 & \textbf{58.57} / 0.83 \\
\cline{2-6}          & 17    & 56.35 / 0.17  & 60.38 / 0.17  & \textbf{45.38} / 0.17 & 46.68 / 0.17 \\
    \hline
    \multicolumn{2}{|c|}{\textbf{Average}} & 54.42 / 0.39  & 51.86 / 0.40  & \textbf{44.49} / 0.37 & 49.22 / 0.36 \\
    \hline

\multicolumn{6}{c}{1: \textit{Traffic} 2:\textit{PeopleOnStreet} 3: \textit{BQTerrace} 4: \textit{Cactus} 5: \textit{BasketballDrive}} \\

\multicolumn{6}{c}{ 6: \textit{ParkScene} 7: \textit{Kimono} 8: \textit{BQMall} 9: \textit{RaceHorses} 10: \textit{Keiba} 11: \textit{Mobisode}} \\

\multicolumn{6}{c}{  12: \textit{Johnny}  13: \textit{FourPeople} 14: \textit{KristenAndSara} 15: \textit{Vidyo1} 16: \textit{Vidyo3} 17: \textit{Vidyo4}} \\

    \end{tabular}%
    \vspace{-3em}
  \label{tab:sub}%
\end{table}%

\textbf{Rate-distortion performance.} We further evaluate the rate-distortion performance of our QE-CNN method in terms of Bjontegaard Distortion-rate (BD-rate) savings over HM 16.0 baseline. In our experiments, the BD-rate is calculated according to the Y-PSNR results and bit-rates at both QP = 22, 27, 32, 37 (Table \ref{tab:BD}-(a)) and QP = 32, 37, 42 and 47 (Table \ref{tab:BD}-(b)). As shown in Table \ref{tab:BD}-(a), our QE-CNN method, which applies QE-CNN-I on I frames and QE-CNN-P on P frames, is able to save BD-rate up to $14.88\%$. Moreover, the average BD-rate saving is $11.06\%$ when applying our QE-CNN method. This BD-rate saving of our QE-CNN method is considerably better than that of AR-CNN and DCAD, which only has $4.96\%$ and $7.98\%$ BD-rate saving, respectively. Moreover, it is shown in Table \ref{tab:BD}-(b) that our QE-CNN method (8.31\%) is still able to save significantly more bit-rates than AR-CNN ($3.96\%$) and DCAD (4.29\%).


\textbf{Transfer to Random Access (RA) mode.} Moreover, we transfer our QE-CNN approach to HEVC RA sequences to verify the generalization capability. Here, the training and test samples are generated by HM 16.0 with RA mode using the default configuration file \textit{encoder\_randomaccess\_main.cfg}. Our QE-CNN-P model, which has been trained by HEVC LDP samples, is fine-tuned by B frames of RA training sequences to enhance the quality of B frames in test sequences compressed by HEVC RA mode. Note that the QE-CNN-I model is not re-trained and directly utilized for I frames of RA sequences.

When fine-tuning the QE-CNN-P model by B frames, the network architecture of QE-CNN-P (i.e., number of layers, filter size and filter number of each layer) is not changed. The trainable parameters (i.e., the weights and bias of each filter) are initialized as those of the original QE-CNN-P model, which is trained by P frames, and then these parameters are updated for B frames. Consequently, at QP = 37, 42.46\% parameters of weight matrices are changed less than 10\%. The absolute change of the 70.13\% weight parameters is less than 30\%. Similar change can be observed in bias matrices. Note that when training QE-CNN-P model from random initialization, the values of 99.02\% parameters are changed with more than 50\%.

\begin{table}[!t]
  \centering
  \vspace{-3.5em}
  \scriptsize
  \caption{$\Delta$PSNR (dB) of our QE-CNN, AR-CNN \cite{dong2015compression}, VRCNN \cite{dai2017convolutional} and DCAD \cite{Wang2017A} methods for HEVC RA sequences.}
  \vspace{-1em}
    \begin{tabular}{|c|c|c|c|c|}
    \hline
    \multicolumn{1}{|c|}{QP} & \multicolumn{2}{c|}{32} & \multicolumn{2}{c|}{37} \\
    \hline
          & I frames & B frames & I frames & B frames \\
    \hline
    AR-CNN \cite{dong2015compression}& 0.1566  & 0.1121  & 0.1964  & 0.1563  \\
    \hline
    VRCNN \cite{dai2017convolutional}& 0.2652  & \multicolumn{1}{c|}{-} & 0.2259  & \multicolumn{1}{c|}{-} \\
    \hline
    DCAD  \cite{Wang2017A} & 0.2100  & 0.1694  & 0.2266  & 0.1910  \\
    \hline
    QE-CNN & \textbf{0.3209} & \textbf{0.3065} & \textbf{0.3522} & \textbf{0.3427} \\
    \hline
    \end{tabular}%
  \label{tab:ra}%
\end{table}%

The results of RA mode are shown in Table \ref{tab:ra}. As shown in Table \ref{tab:ra}, our QE-CNN method still obviously outperforms the conventional AR-CNN, VRCNN and DCAD methods on both I and B frames. For example, the $\Delta$PSNR of our QE-CNN method (0.3065 dB) is almost twice higher that of DCAD (0.1694 dB) and three times higher than AR-CNN (0.1121 dB) for B frames at QP = 32. Besides, the $\Delta$PSNR results of RA are comparable to those of LDP on both I frames and inter-coding frames (P frames in LDP and B frames in RA). This validates the effectiveness of our QE-CNN method for RA configuration.

\vspace{-.5em}
\subsection{Ablation experiments}\label{abl}

In addition, we conduct the ablation experiments on test sequences to evaluate the quality enhancement performance of our QE-CNN model when adding more layers or changing the activation function. These results validate that our QE-CNN structure achieves the optimal performance.
\begin{figure*}[!t]
\centering
\vspace{-3.5em}
\subfigure{\includegraphics[width = 0.32\linewidth]{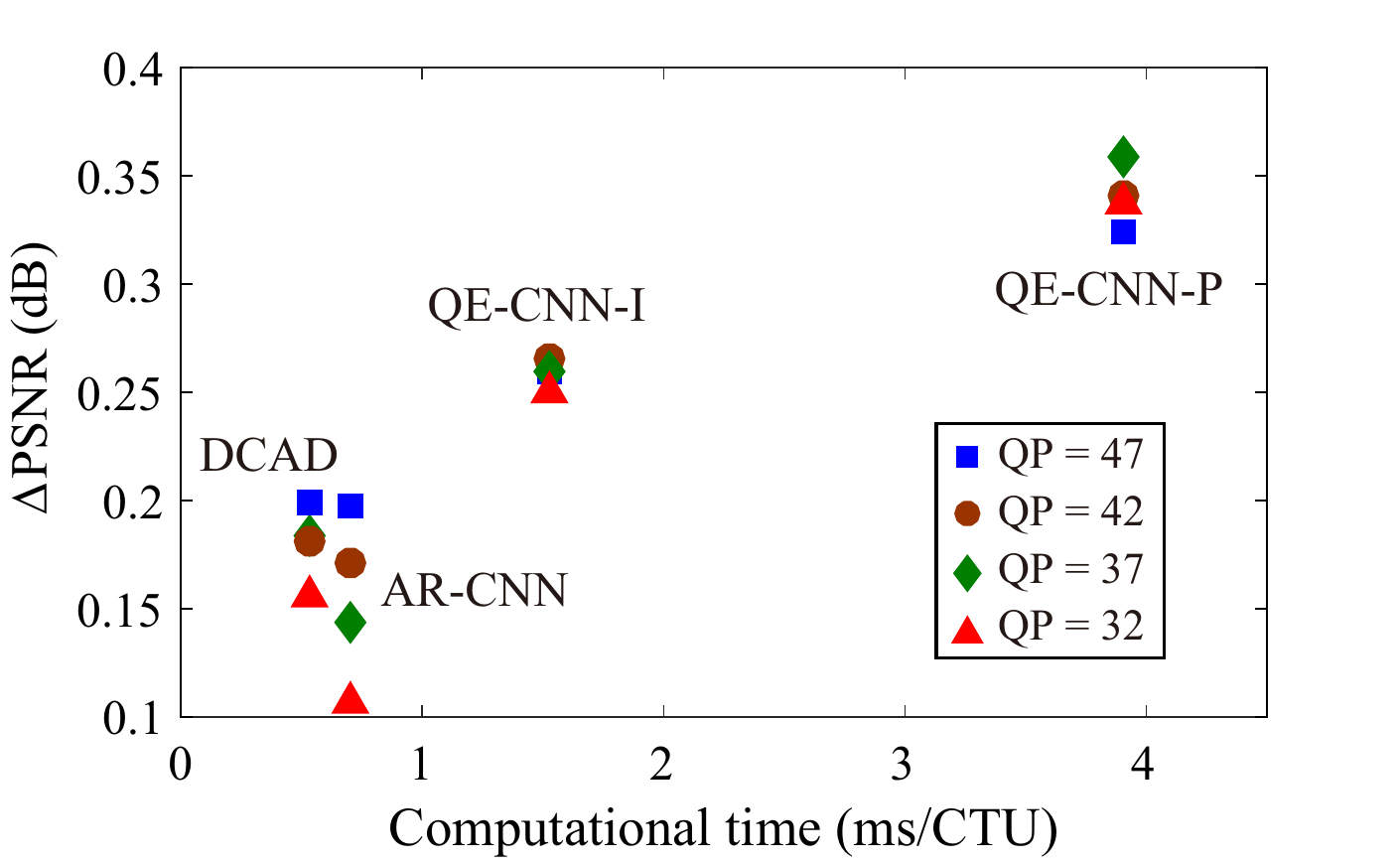}}
\subfigure{\includegraphics[width = 0.32\linewidth]{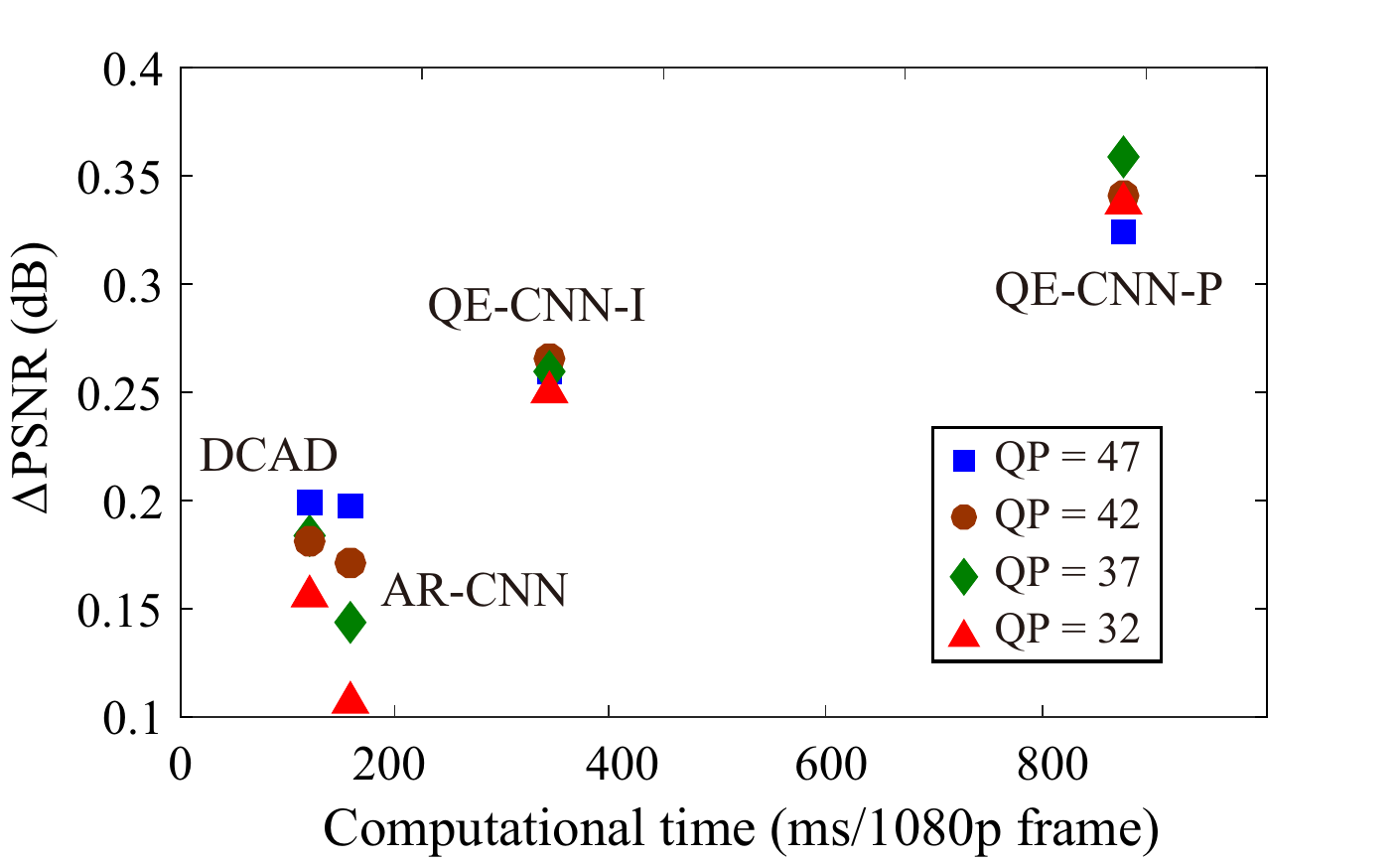}}
\subfigure{\includegraphics[width = 0.32\linewidth]{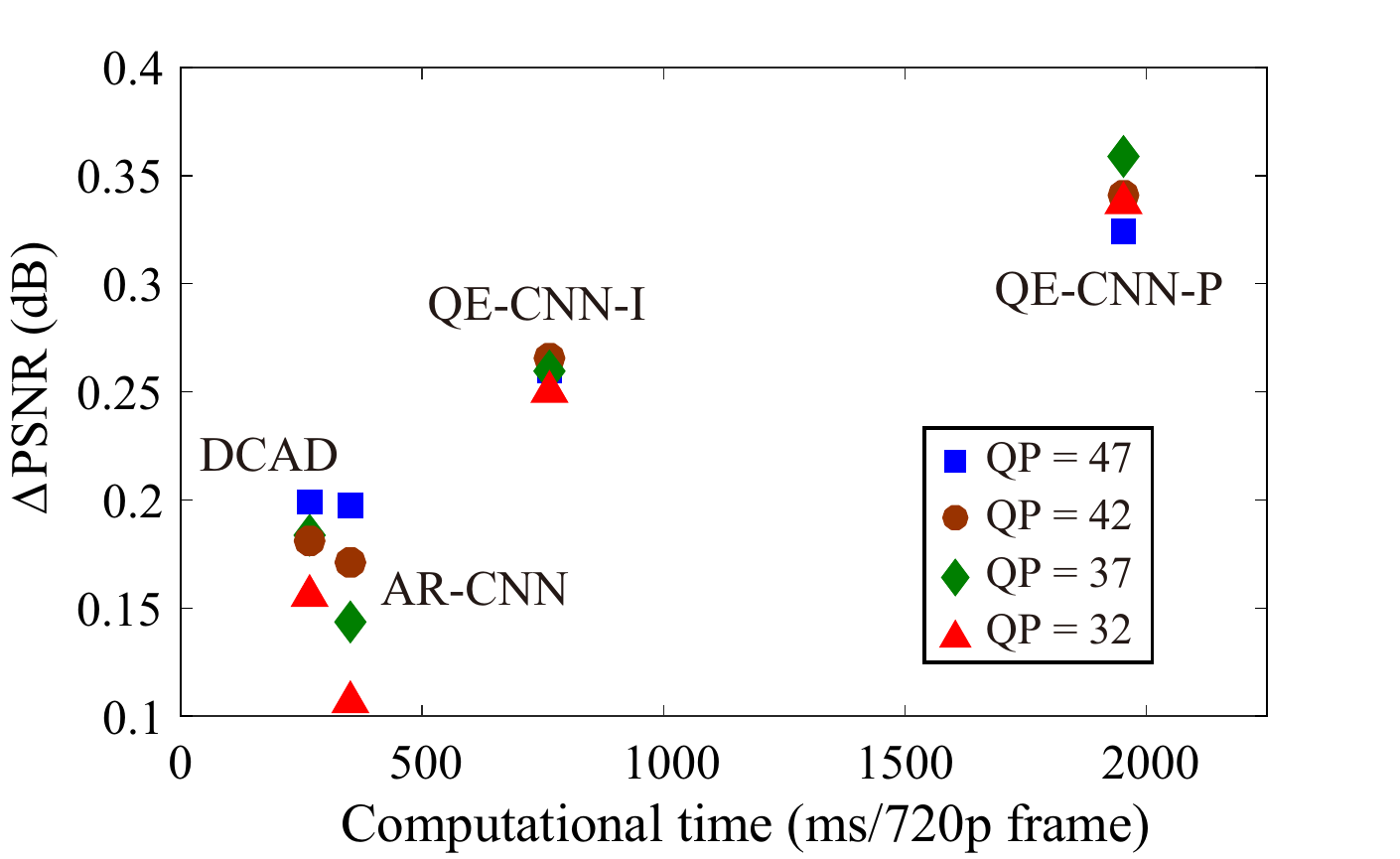}}
\vspace{-1em}
\caption{Computational time versus quality enhancement for our QE-CNN, AR-CNN \cite{dong2015compression} and DCAD \cite{Wang2017A} methods on HEVC LDP videos.}\label{time}
\vspace{-1.5em}
\end{figure*}

\begin{table}[!t]
  \centering
  \scriptsize
  \vspace{-1.5em}
  \caption{$\Delta$PSNR (dB) of QE-CNN with additional layers at QP = 42.}
  \vspace{-1em}
    \begin{tabular}{|c|c|c|c|}
    \hline
          & \multicolumn{2}{c|}{QE-CNN-I} & QE-CNN-P \\
    \hline
          & I frames & P frames & P frames  \\
    \hline
    Original& \textbf{0.3469} & 0.2650  & \textbf{0.3407}   \\
    \hline
    Adding 1 layer&  0.3331  & 0.2287  & 0.3358   \\
    \hline
    Adding  2 layers& 0.3100  & 0.2153  & 0.3315  \\
    \hline
    \end{tabular}%
     \vspace{-2.5em}
  \label{tab:morelayer}%
\end{table}%

\textbf{Adding more layers.} We first measure the $\Delta$PSNR results on the test sequences, after adding more layers to our QE-CNN method. Table VIII shows the $\Delta$PSNR results of adding more layers into QE-CNN for the scenario of QP = 42. Specifically, we insert one layer (64 filters with size of 3$\times$3) between Conv 3 and Conv 4 for QE-CNN-I. We can see that $\Delta$PSNR of QE-CNN-I decreases from 0.3469 dB to 0.3331 dB on I frames, and from 0.2650 dB to 0.2153 dB for P frames. Given this QE-CNN-I model with one additional layer, we further add one layer (64 filters with size of 3$\times$3) between Conv 7 and Conv 9 for QE-CNN-P. This results in the decrease of $\Delta$PSNR from 0.3407 dB to 0.3358 dB on P frames. Moreover, two additional layers between Conv 3 and Conv 4 for QE-CNN-I and Conv 7 and Conv 9 for QE-CNN-P lead to further degradation of $\Delta$PSNR. To summarize, adding more layers to QE-CNN decreases the performance of our method.

\begin{table}[!t]
  \centering
  \scriptsize
  \caption{$\Delta$PSNR (dB) of QE-CNN with various activation functions.}
  \vspace{-1em}
    \begin{tabular}{|c|c|c|c|c|}
    \hline
     QP & \multicolumn{2}{c|}{42} & \multicolumn{2}{c|}{32} \\
    \hline
          & I frames & P frames & I frames & P frames \\
    \hline
    DS-CNN \cite{Yang2017coding}& 0.3165  & 0.3165  & 0.3172  & 0.3018  \\
    \hline
    QE-CNN (ReLU) & 0.3165  & 0.3002  & 0.3172  & 0.3186  \\
    \hline
    QE-CNN (Leaky-ReLU) & 0.3146  & 0.3370 & 0.2711  & 0.3352 \\
    \hline
    QE-CNN (PReLU) & \textbf{0.3469} & \textbf{0.3407} & \textbf{0.3323} & \textbf{0.3367} \\
    \hline
    \end{tabular}%
    \vspace{-2.5em}
  \label{tab:ra}%
\end{table}%
\textbf{Comparison between ReLU, Leaky-ReLU and PReLU.} Table \ref{tab:ra} also shows the $\Delta$PSNR test results of our QE-CNN method with ReLU, Leaky-ReLU and PReLU. It can be seen from Table IX that PReLU achieves the best performance among the three activation functions for quality enhancement. For example, when using PReLU, averaged PSNR can be improved by 0.3469 dB and 0.3407 dB on I and P frames at QP = 42, respectively. Such performance decreases to 0.3165 dB for I frames and 0.3002 dB for P frames, after replacing PReLU by ReLU. For QE-CNN with Leaky-ReLU, $\Delta$PSNR = 0.3146 dB and 0.3370 dB for I and P frames, respectively. This validates applying PReLU in our QE-CNN model outperforms using ReLU or Leaky-ReLU.

\textbf{Comparison between DS-CNN \cite{Yang2017coding} and QE-CNN.} Moreover, we compare the performance of our QE-CNN method with DS-CNN of our conference paper \cite{Yang2017coding}. The $\Delta$PSNR results of QE-CNN and DS-CNN are reported in Table \ref{tab:ra}. As Table \ref{tab:ra} shows, for our QE-CNN model, $\Delta$PSNR increases from 0.3165 dB to 0.3469 dB for I frames at QP = 42, in comparison with DS-CNN \cite{Yang2017coding}. The $\Delta$PSNR results of QE-CNN are also improved from 0.3162 to 0.3407 dB for P frames at QP = 42, compared with DS-CNN \cite{Yang2017coding}. Similar results can be found from Table \ref{tab:ra} for QP = 37. Hence, our QE-CNN method outperforms the previous DS-CNN \cite{Yang2017coding} for HEVC quality enhancement.

%
\vspace{-1em}
\subsection{Computational time analysis}\label{ti}

Moreover, we evaluate the computational time of our QE-CNN method for quality enhancement compared to the conventional AR-CNN \cite{dong2015compression}. In our experiments, we evaluate the computational time via the running time of quality enhancement on an Ubuntu PC equipped with one GeForce GTX 1080 GPU. Fig. \ref{time} shows the computational time of our QE-CNN method and the conventional AR-CNN and DCAD methods, as well as their Y-PSNR improvement. Note that the results in this figure are obtained by averaging over all the test sequences.

As shown in Fig. \ref{time}, our QE-CNN method performs considerably better than the conventional AR-CNN and DCAD methods in terms of Y-PSNR improvement. However, this improvement occurs at the expense of computational time. Specifically, the running time of AR-CNN method is 0.70 ms per Coding Tree Unit (CTU) and that of DCAD is 0.64 ms per CTU. In contrast, our QE-CNN-I model requires approximately 1.53 ms per CTU, and QE-CNN-P consumes 3.90 ms per CTU. Thus, the performance improvement of our QE-CNN method is at the expense of computational time. The next section primarily focuses on optimizing quality enhancement of our QE-CNN method under time constraint, and a prototype for real-time implementation is designed.

\section{Time-constrained quality enhancement optimization}\label{TQEO}

As discussed in Section \ref{ti}, our QE-CNN method may consume large amounts of computational time for enhancing the quality of HEVC videos, especially for high-resolution videos or with insufficient computational resources. Meanwhile, in many application scenarios, e.g., real-time video play, quality enhancement has to be constrained by a time limitation. Hence, we propose the TQEO scheme, i.e., time-constrained quality enhancement optimization scheme, enabling our QE-CNN method to be practical for time-constrained scenarios.

\vspace{-.5em}
\subsection{Formulation for TQEO scheme}

\begin{figure}[!t]
\centering
\vspace{-3em}
\hspace{-1.5em}\includegraphics[width = 1\linewidth]{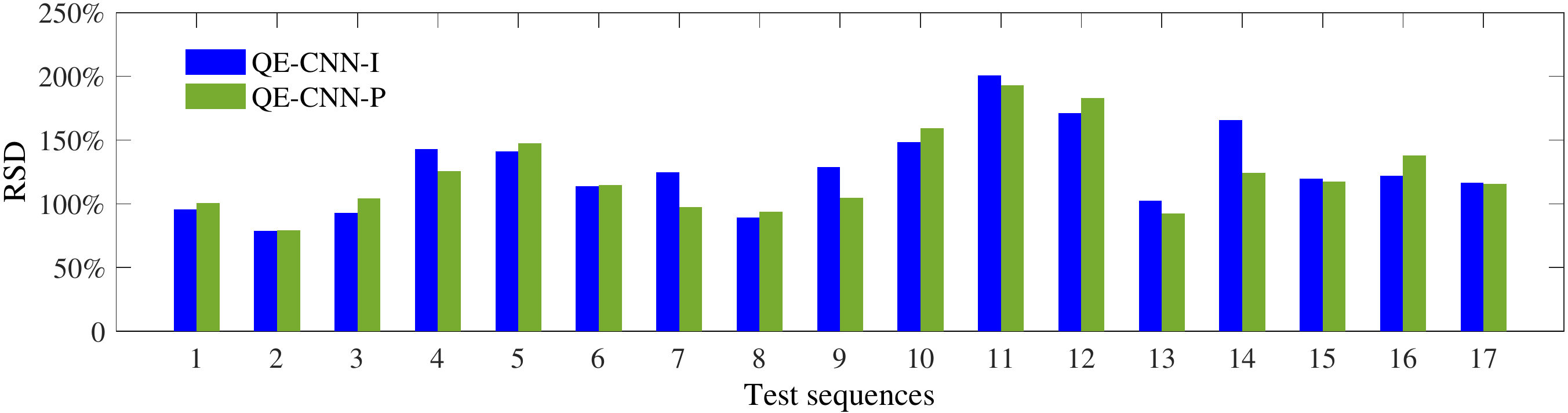}\\
\scriptsize{1: \textit{Traffic} 2: \textit{PeopleOnStreet} 3: \textit{BQTerrace} 4: \textit{Cactus} 5: \textit{BasketballDrive} 6: \textit{ParkScene}\\ 7: \textit{Kimono} 8: \textit{BQMall}  9: \textit{RaceHorses} 10: \textit{Keiba} 11: \textit{Mobisode} 12: \textit{Johnny} \\13: \textit{FourPeople} 14: \textit{KristenAndSara} 15: \textit{Vidyo1} 16: \textit{Vidyo3} 17: \textit{Vidyo4} }
\vspace{-1em}
\caption{RSD of the MSE reduction across CTUs in each test sequence. For each sequence, the RSD of the MSE reduction is averaged over all frames. Note that all test sequences are compressed by HEVC LDP mode at QP = 42, and then QE-CNN is applied for quality enhancement.}\label{RSD}
\vspace{-1.5em}
\end{figure}

To meet the time constraint, the computational time of our QE-CNN method needs to be reduced. The computational time can be reduced by applying QE-CNN to only some of the CTUs at a frame. Interestingly, we find that the quality enhanced by our QE-CNN method varies across different CTUs. To quantify this variation, we evaluate the Relative Standard Deviations (RSDs) of the MSE reduction among CTUs at a frame when applying our QE-CNN method. Note that RSD is the ratio of standard deviation divided by the corresponding mean value. For each sequence, the RSD of the MSE reduction is averaged over all frames. As shown in Fig. \ref{RSD}, the RSD of the MSE reduction is approximately or even greater than $100\%$ for all 17 test sequences. In other words, the standard deviation is comparable to its corresponding mean value. This result indicates a large difference in quality enhancement across different CTUs. Therefore, under the time constraint, our QE-CNN method needs to be conducted on CTUs that may achieve large quality enhancement, in order to maximize the quality enhancement.

Our TQEO scheme aims to control the computational time of our QE-CNN method through a constraint while optimizing quality enhancement. The formulation of our TQEO scheme can be modeled as
\begin{eqnarray}
\label{formu}
\max_{\{k_n\}_{n=1}^N} \sum_{n=1}^N \Delta\text{MSE}_n(k_n)
\quad \text{s.t.} \quad \sum_{n=1}^N t_n(k_n) \leq T,
\end{eqnarray}
where $\Delta\text{MSE}_n(k_n)$ is defined as the MSE reduction of the $n$-th CTU and $t_n(k_n)$ is the computational time for quality enhancement on the $n$-th CTU. Here, $k_n \in \{0,1,2\}$ is a variable to be calculated for solving \eqref{formu}. In particular, $k_n=0$ means that the $n$-th CTU is without any quality enhancement, $k_n=1$ stands for applying QE-CNN-I in the CTU, and $k_n=2$ indicates applying QE-CNN-P. Additionally, $T$ denotes the time constraint, and $N$ is the total number of CTUs at a frame. Next, we discuss the solution to our TQEO formulation for I and P frames, respectively.

\vspace{-.5em}
\subsection{Solution to \eqref{formu} for I frames} \label{sol_I}

To solve \eqref{formu} for I frames, we first model $t_n(k_n)$ and $\Delta\text{MSE}_n(k_n)$ by training on 10 sequences randomly selected from our dataset introduced in Section \ref{CNNP}. Note that these 10 training sequences do not overlap with all the 17 test sequences of Sections \ref{EX-QE} and \ref{EX-TC}. Recall that only the network of QE-CNN-I can be used on I frames; thus, $k_n \in \{0,1\}$ for I frames. Clearly, we have $t_n(k_n=1) = 0$ when the CTU is without any quality enhancement. For modeling $t_n(k_n=1)$, we record the computational time of QE-CNN-I among all CTUs of the training sequences when running on a computer with one GeForce GTX 1080 GPU. Consequently, the mean and standard deviation of $t_n(k_n=1)$ are $1.536$ ms and $0.096$ ms, respectively. Since the standard deviation is significantly smaller than the mean value for $t_n(k_n=1)$, we simply set $t_n(k_n=1)$ as a constant $1.536 \ \text{ms}$ in our TQEO scheme. Thus, given the constraint of $\sum_{n=1}^N t_n(k_n) \leq T$ in \eqref{formu}, we can obtain that the number of CTUs with $k_n=1$ at a frame is equivalent to
\begin{eqnarray}
\label{N1}
N_1 = \lfloor {T}/{t_n(k_n=1)} \rfloor,
\end{eqnarray}
where $\lfloor\cdot\rfloor$ represents rounding down. In the following, we model $\Delta\text{MSE}_n(k_n)$ to solve the optimization problem of $\max\sum_{n=1}^N \Delta\text{MSE}_n(k_n)$ for \eqref{formu}.

\begin{figure}[!t]
\centering
\vspace{-3em}
\subfigure{\includegraphics[width = 0.49\linewidth]{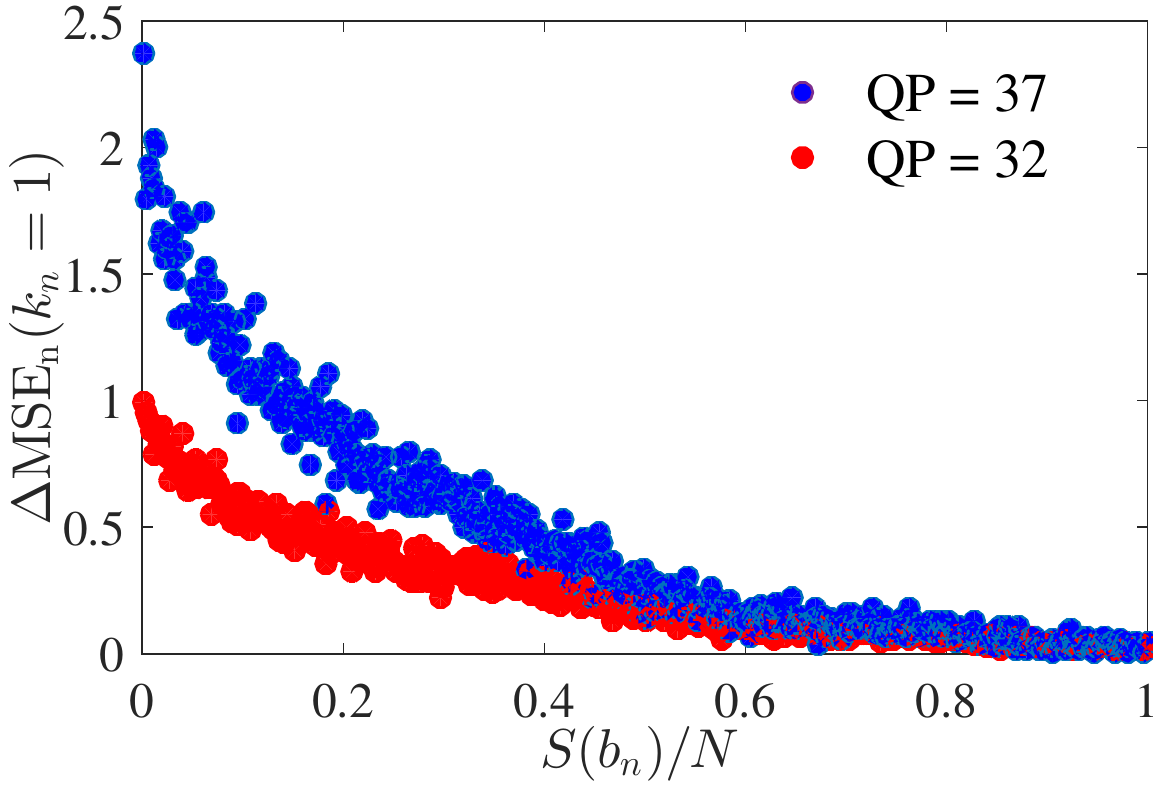}}
\subfigure{\includegraphics[width = 0.49\linewidth]{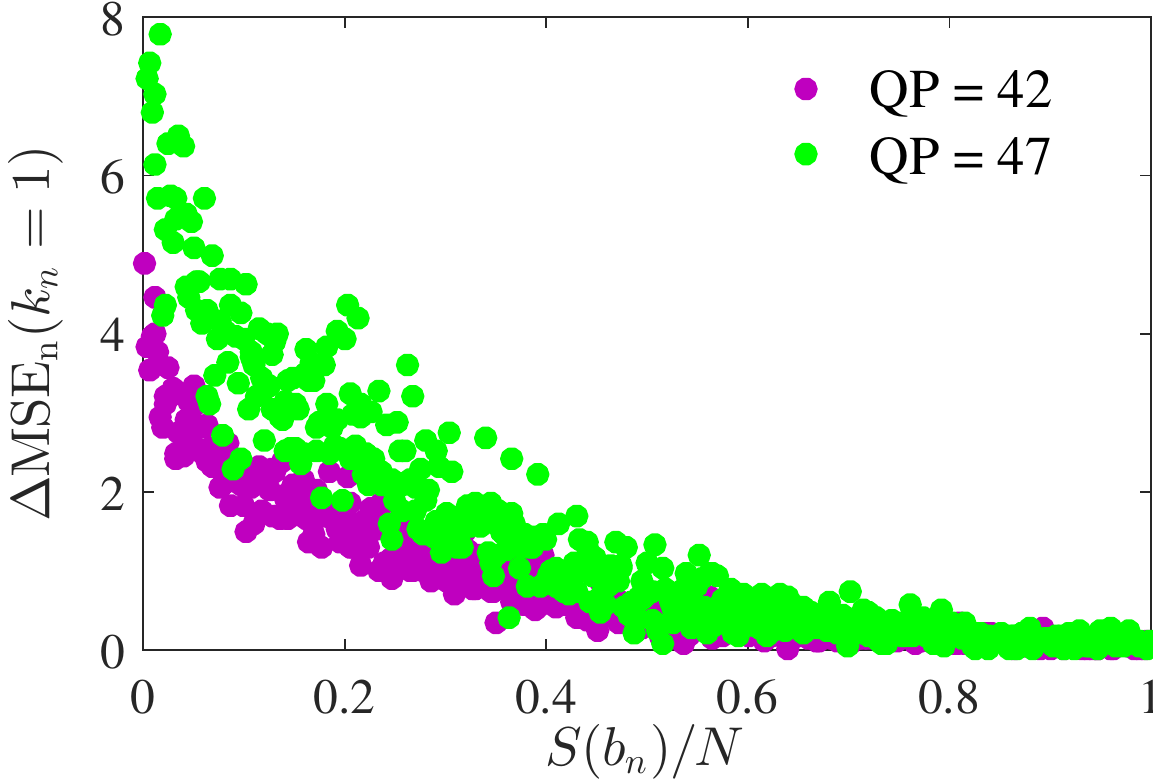}}
\vspace{-2em}
\caption{Relationship between $\Delta\text{MSE}_n(k_n)$ and $S(b_n)$ for I frames. Note that $\Delta\text{MSE}_n(k_n)$ for each $S(b_n)$ is averaged among all training sequences.}\label{train_I}
\vspace{-1em}
\end{figure}

For I frames, we find that $\Delta\text{MSE}_n(k_n)$ of different CTUs has a strong correlation with their corresponding bit allocation in HEVC, denoted by $b_n$. The Spearman correlation coefficients between $\Delta\text{MSE}_n(k_n)$ and $b_n$, averaged over all training sequences, are 0.85, 0.80, 0.73 and 0.65 for QP = 32, 37, 42 and 47, respectively. Additionally, as shown in Fig. \ref{train_I}, $S(b_n)$ is denoted as the rank of descending sorted $\{b_n\}_{n=1}^N$ in a frame\footnote{Descending sort consumes only 0.020 ms per 1080p frame by an Intel Core(TM) i7-4790K CPU, which can be ignored in our TQEO scheme.}, and $\Delta\text{MSE}_n(k_n)$ reduces along with the descending $b_n$. Therefore, at I frames, CTUs with more allocated bits are more likely to achieve higher quality enhancement, and they should be with high priority in enhancing quality under the time constraint. Thus, given $N_1 = \lfloor {T}/{t_n(k_n=1)} \rfloor $, the solution to \eqref{formu} can be obtained as
\begin{equation}
k_n=\left\{
\begin{aligned}
1&,&\quad S(b_n) \leq N_1,  \\
0&,&\quad S(b_n) > N_1,
\end{aligned}
\right.
\end{equation}
for I frames of HEVC compressed videos.

\begin{table}[t]
  \centering
  \scriptsize
  \caption{Coefficients in the function of $\Delta\text{MSE}_n(k_n)$.}
  \vspace{-1em}
    \begin{tabular}{|c|c|c|c|c|}
    \hline
    QP & 32 &  37 &  42 &  47 \\
    \hline
    \hline
    $a_1$ & 0.643 & 0.429 & 3.693 & 10.85 \\
    \hline
    $b_1$ & 2.672 & 2.841 & 9.728 & 22.71 \\
    \hline
    $c_1$ & 2.061 & 2.344 & 6.266 & 12.34 \\
    \hline
    \textbf{R-Square} &  \textbf{0.995}     &  \textbf{0.994}     & \textbf{0.995}      &\textbf{0.990}  \\
    \hline
    \hline
    $a_2$ & 1.218 & 1.476 & 8.928 & 21.64 \\
    \hline
    $b_2$ & 4.352 & 4.588 & 20.54 & 42.00 \\
    \hline
    $c_2$ & 3.177 & 3.265 & 12.08 & 21.50 \\
    \hline
    \textbf{R-Square} & \textbf{0.997}      &  \textbf{0.996}     & \textbf{0.996}      & \textbf{0.990} \\
    \hline
    \end{tabular}%
    \vspace{-1.5em}
  \label{coe}%
\end{table}%

\begin{table}[!t]
  \scriptsize
  \centering
  \caption{Solution to \eqref{formu2} for 1080p sequences.}
  \vspace{-1em}
    \begin{tabular}{|c|c|c|c|c|c|c|c|c|}
    \hline
    \multirow{2}[4]{*}{$T/T_{\max}$} & \multicolumn{2}{c|}{QP = 32} & \multicolumn{2}{c|}{QP = 37} & \multicolumn{2}{c|}{QP = 42} & \multicolumn{2}{c|}{QP = 47} \\
\cline{2-9}          & $N1$  & $N2$  & $N1$  & $N2$  & $N1$  & $N2$  & $N1$  & $N2$ \\
    \hline
    $10\%$ & 121   & 0     & 121   & 0     & 119   & 1     & 119   & 1 \\
    \hline
    $20\%$ & 241   & 1     & 243   & 0     & 132   & 44    & 165   & 31 \\
    \hline
    $30\%$ & 259   & 42    & 325   & 16    & 132   & 92    & 142   & 88 \\
    \hline
    $40\%$ & 231   & 101   & 292   & 77   & 132   & 140   & 132   & 140 \\
    \hline
    $50\%$ & 203   & 160   & 246   & 143   & 114   & 195   & 132   & 188 \\
    \hline
    $60\%$ & 170   & 221   & 198   & 210   & 109   & 245   & 114   & 243 \\
    \hline
    $70\%$ & 137   & 282   & 142   & 280   & 99    & 297   & 109   & 293 \\
    \hline
    $80\%$ & 99    & 345   & 81    & 352   & 86    & 350   & 109   & 341 \\
    \hline
    $90\%$ & 58    & 409   & 0     & 432   & 66    & 406   & 76    & 402 \\
    \hline
    \end{tabular}%
  \label{tab:solution}%
  \vspace{-2.5em}
\end{table}%

\vspace{-1em}
\subsection{Solution to \eqref{formu} for P frames} \label{sol_P}

To solve \eqref{formu} for P frames, we model $t_n(k_n)$ and $\Delta\text{MSE}_n(k_n)$ of \eqref{formu} as follows. First, the computational time of QE-CNN-I on P frames is the same as that on I frames; thus, $t_n(k_n=1) = 1.536 \ \text{ms}$ still holds for P frames. Similarly, QE-CNN-P consumes $3.900\ \text{ms}$ per CTU on average, with a small deviation of $0.120$ ms. Thus, $t_n(k_n=2)$ is also set by a constant, i.e., $3.900\ \text{ms}$.

\begin{figure*}[!t]
\centering
\vspace{-3em}
\subfigure[QP = 32]{\includegraphics[width = 0.245\linewidth]{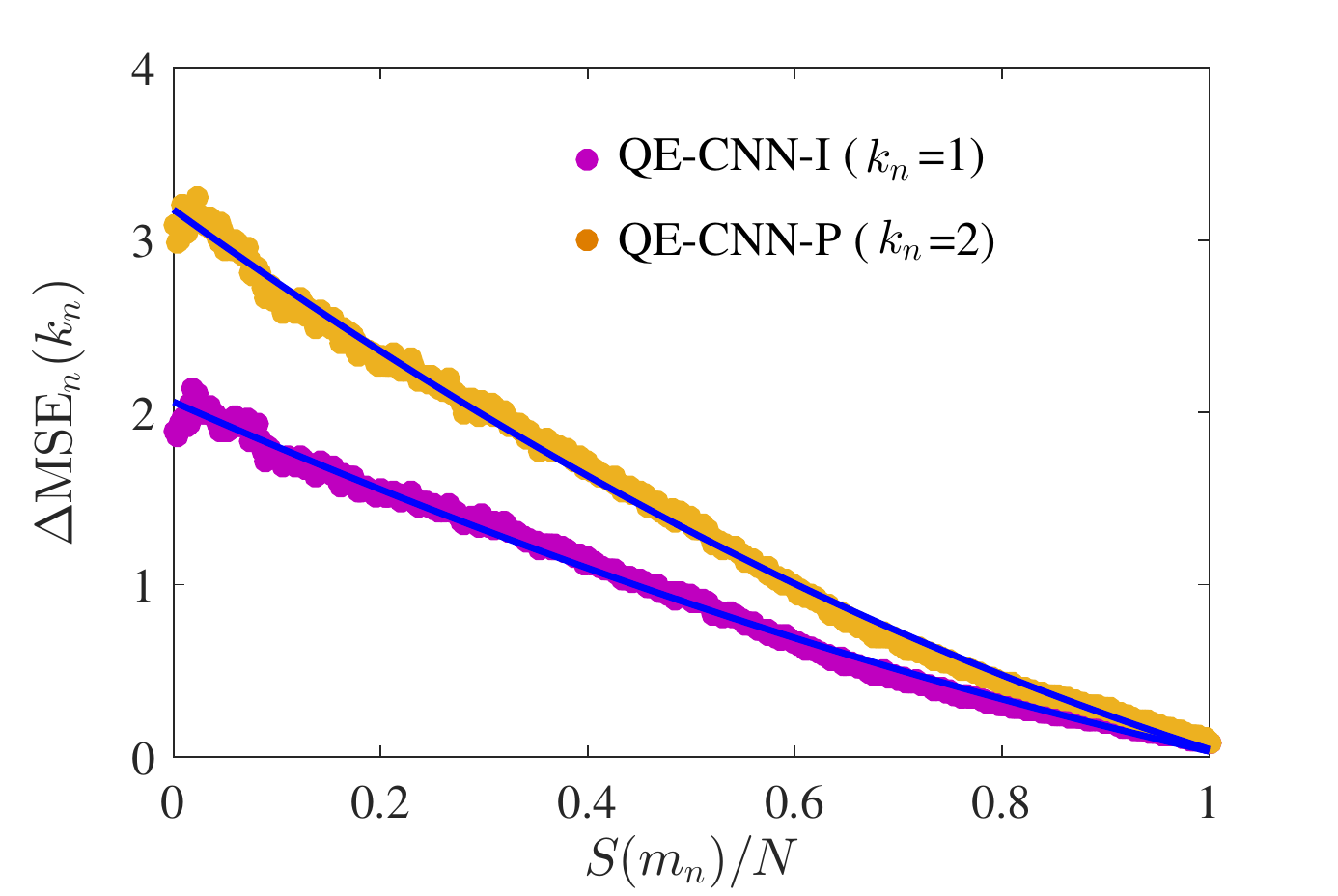}}
\subfigure[QP = 37]{\includegraphics[width = 0.245\linewidth]{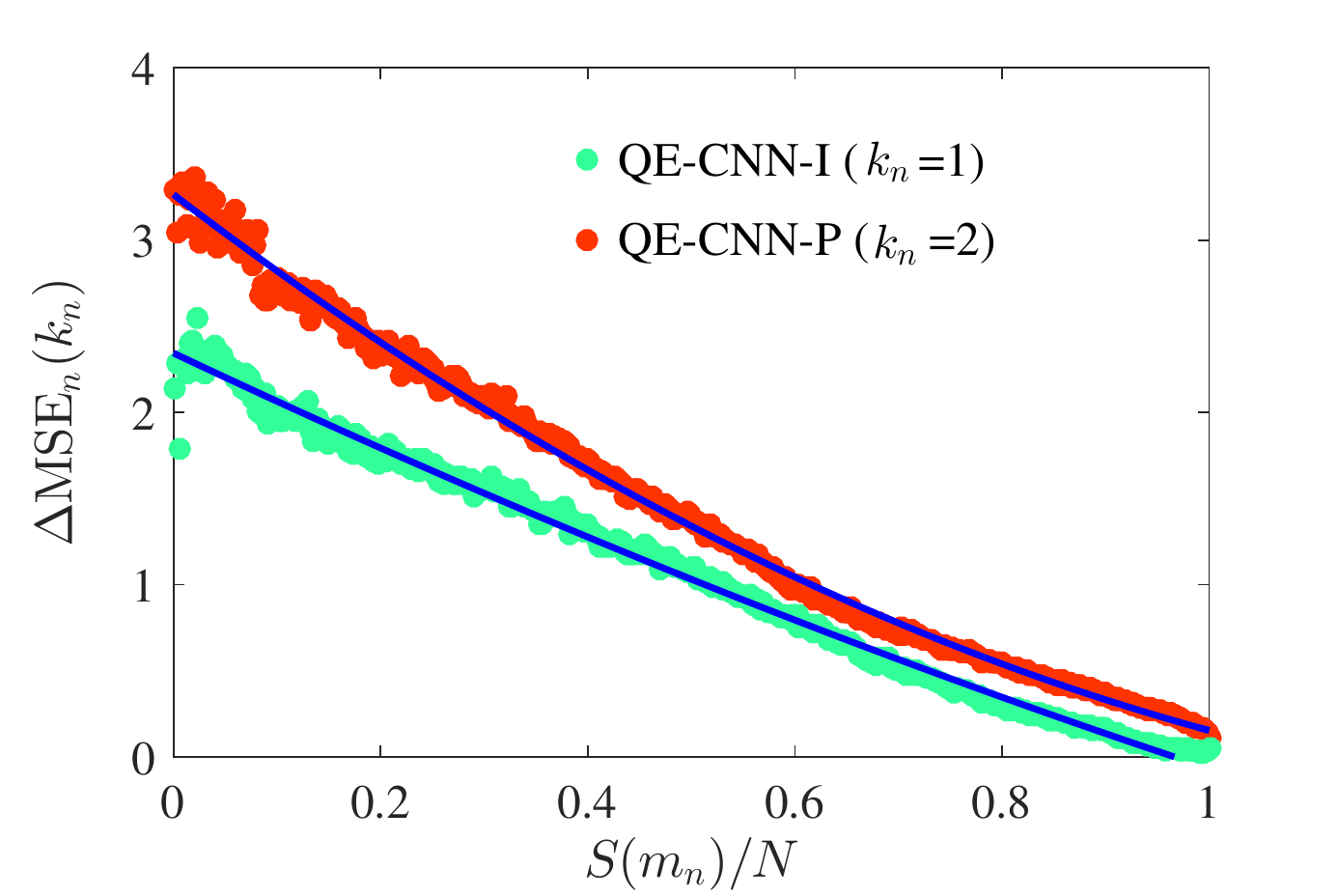}}
\subfigure[QP = 42]{\includegraphics[width = 0.245\linewidth]{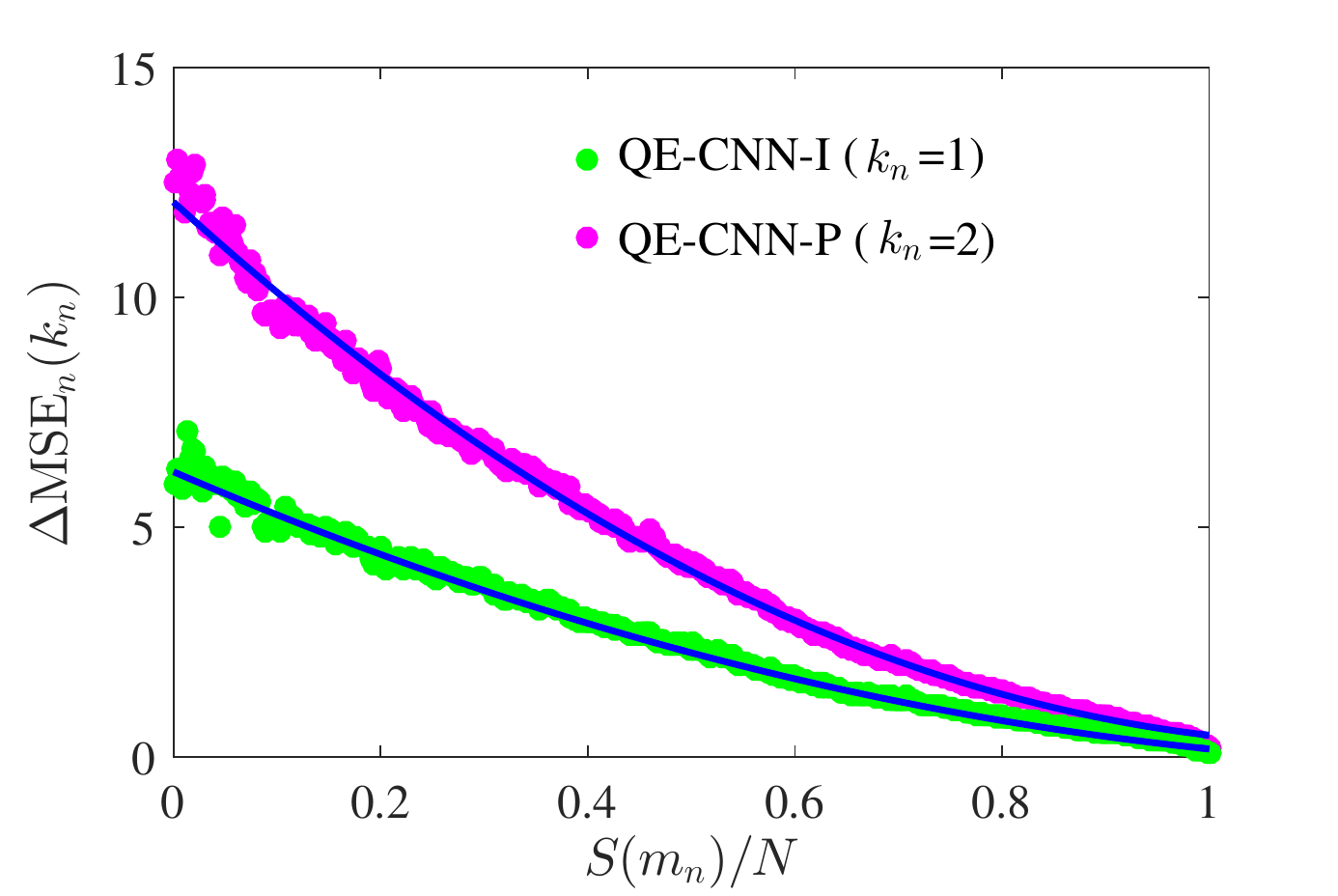}}
\subfigure[QP = 47]{\includegraphics[width = 0.245\linewidth]{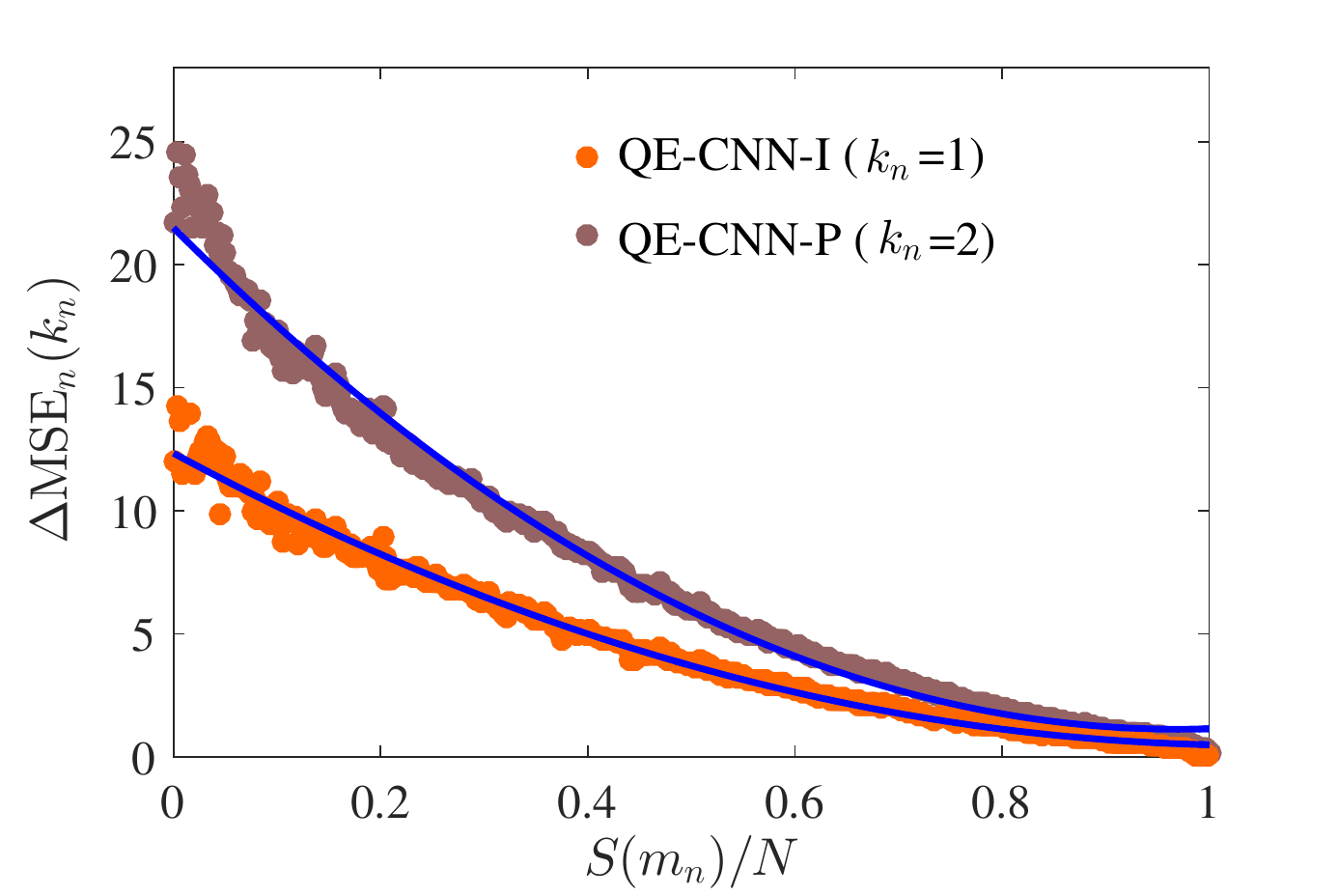}}
\vspace{-1.5em}
\caption{Relationship of $\Delta\text{MSE}_n(k_n)$ for P frames. The points of $\Delta\text{MSE}_n(k_n)$ for each $S(m_n)$ are the average values among the 10 training sequences.}\label{train_P}
\end{figure*}

\begin{table*}[!t]
  \centering
  \vspace{-1.5em}
  \scriptsize
  \caption{Results of time control accuracy for our TQEO scheme.}
  \vspace{-1em}
    \begin{tabular}{|p{.7cm}<{\centering}|c|c|c|c|c|c|c|c|c|c|c|c|c|c|c|c|c|}
    \hline
    \multicolumn{2}{|c|}{$T/T_{\max} (\%)$} &
      \multicolumn{4}{c|}{QP = 32} &
      \multicolumn{4}{c|}{QP = 37} &
      \multicolumn{4}{c|}{QP = 42} &
      \multicolumn{4}{c|}{QP = 47}
      \\
    \hline
    Class & Seq. &
      20 &
      40 &
      60 &
      80 &
      20 &
      40 &
      60 &
      80 &
      20 &
      40 &
      60 &
      80 &
      20 &
      40 &
      60 &
      80
      \\
    \hline
    \multirow{2}[2]{*}{A} &
      1 &
      20.11 &
      40.20 &
      60.16 &
      80.08 &
      20.10 &
      40.24 &
      60.21 &
      80.10 &
      19.04 &
      39.99 &
      60.00 &
      79.99 &
      19.66 &
      40.16 &
      60.12 &
      80.13
      \\
\cline{2-18}     &
      2 &
      20.12 &
      40.12 &
      60.09 &
      80.05 &
      20.08 &
      40.23 &
      60.22 &
      80.09 &
      19.93 &
      39.97 &
      59.99 &
      79.98 &
      20.09 &
      40.11 &
      60.21 &
      80.09
      \\
    \hline
    \multirow{5}[2]{*}{B} &
      3 &
      20.21 &
      40.18 &
      60.15 &
      80.08 &
      20.16 &
      40.31 &
      60.25 &
      80.11 &
      19.93 &
      39.95 &
      60.00 &
      79.98 &
      20.11 &
      40.09 &
      60.02 &
      79.97
      \\
\cline{2-18}     &
      4 &
      20.18 &
      40.18 &
      60.15 &
      80.09 &
      20.16 &
      40.22 &
      60.16 &
      80.04 &
      19.88 &
      39.97 &
      59.95 &
      79.95 &
      20.12 &
      40.12 &
      60.10 &
      80.12
      \\
\cline{2-18}     &
      5 &
      20.17 &
      40.20 &
      60.13 &
      80.09 &
      20.15 &
      40.20 &
      60.18 &
      80.04 &
      19.92 &
      39.96 &
      59.98 &
      79.90 &
      20.13 &
      40.15 &
      60.10 &
      80.06
      \\
\cline{2-18}     &
      6 &
      20.14 &
      40.20 &
      60.14 &
      80.09 &
      20.12 &
      40.23 &
      60.18 &
      80.08 &
      19.95 &
      39.98 &
      59.98 &
      80.00 &
      20.09 &
      40.13 &
      60.07 &
      80.11
      \\
\cline{2-18}     &
      7 &
      20.15 &
      40.20 &
      60.17 &
      80.12 &
      20.14 &
      40.23 &
      60.16 &
      80.04 &
      19.95 &
      39.97 &
      59.97 &
      79.91 &
      20.13 &
      40.11 &
      60.06 &
      80.03
      \\
    \hline
    \multirow{4}[2]{*}{C} &
      8 &
      20.25 &
      40.26 &
      60.25 &
      80.14 &
      20.20 &
      40.09 &
      60.19 &
      80.04 &
      19.53 &
      39.91 &
      59.90 &
      79.93 &
      20.23 &
      40.44 &
      60.31 &
      80.06
      \\
\cline{2-18}     &
      9 &
      20.15 &
      40.00 &
      60.01 &
      79.98 &
      20.26 &
      39.98 &
      60.00 &
      79.78 &
      18.11 &
      39.29 &
      59.91 &
      79.95 &
      18.87 &
      38.89 &
      59.97 &
      80.08
      \\
\cline{2-18}     &
      10 &
      20.23 &
      40.10 &
      60.22 &
      80.04 &
      20.29 &
      40.09 &
      60.20 &
      80.01 &
      19.24 &
      39.94 &
      59.92 &
      79.92 &
      19.72 &
      40.41 &
      60.32 &
      80.08
      \\
\cline{2-18}     &
      11 &
      20.16 &
      40.08 &
      60.14 &
      79.98 &
      20.21 &
      40.13 &
      60.22 &
      80.00 &
      19.20 &
      39.89 &
      59.89 &
      79.93 &
      19.70 &
      40.51 &
      60.34 &
      80.14
      \\
    \hline
    \multirow{3}[2]{*}{E} &
      12 &
      20.11 &
      40.11 &
      60.14 &
      80.04 &
      20.06 &
      40.22 &
      60.15 &
      80.08 &
      19.96 &
      39.99 &
      59.98 &
      79.99 &
      20.07 &
      40.15 &
      60.09 &
      80.07
      \\
\cline{2-18}     &
      13 &
      20.21 &
      40.30 &
      60.40 &
      80.45 &
      20.07 &
      40.22 &
      60.07 &
      80.07 &
      19.94 &
      40.00 &
      60.00 &
      80.03 &
      20.08 &
      40.15 &
      60.11 &
      80.07
      \\
\cline{2-18}     &
      14 &
      20.13 &
      40.15 &
      60.15 &
      80.04 &
      20.02 &
      40.22 &
      60.15 &
      80.10 &
      19.98 &
      40.03 &
      59.99 &
      80.00 &
      20.09 &
      40.14 &
      60.07 &
      80.06
      \\
    \hline
    \multirow{3}[2]{*}{E'} &
      15 &
      19.98 &
      40.06 &
      60.10 &
      80.07 &
      20.00 &
      40.23 &
      60.15 &
      80.11 &
      19.94 &
      40.00 &
      59.98 &
      79.97 &
      19.12 &
      39.88 &
      59.77 &
      79.65
      \\
\cline{2-18}     &
      16 &
      20.08 &
      40.08 &
      60.04 &
      80.01 &
      20.06 &
      40.21 &
      60.08 &
      80.09 &
      19.88 &
      39.97 &
      59.95 &
      80.01 &
      19.64 &
      40.08 &
      60.04 &
      80.01
      \\
\cline{2-18}     &
      17 &
      20.11 &
      40.11 &
      60.10 &
      80.06 &
      20.03 &
      40.22 &
      60.11 &
      80.07 &
      19.95 &
      39.98 &
      59.97 &
      79.98 &
      20.07 &
      40.13 &
      60.06 &
      80.06
      \\
    \hline
    \multicolumn{2}{|c|}{\textbf{Average MAE\ $(\%)$}} &
      \textbf{0.149} &
      \textbf{0.149} &
      \textbf{0.150} &
      \textbf{0.087} &
      \textbf{0.124} &
      \textbf{0.195} &
      \textbf{0.158} &
      \textbf{0.077} &
      \textbf{0.333} &
      \textbf{0.075} &
      \textbf{0.039} &
      \textbf{0.039} &
      \textbf{0.266} &
      \textbf{0.241} &
      \textbf{0.134} &
      \textbf{0.092}
      \\
    \hline
    \multicolumn{18}{c}{1: \textit{Traffic} 2:\textit{PeopleOnStreet} 3: \textit{BQTerrace} 4: \textit{Cactus} 5: \textit{BasketballDrive} 6: \textit{ParkScene} 7: \textit{Kimono} } \\

\multicolumn{18}{c}{8: \textit{BQMall} 9: \textit{RaceHorses} 10: \textit{Keiba} 11: \textit{Mobisode} 12: \textit{Johnny}  13: \textit{FourPeople} 14: \textit{KristenAndSara} 15: \textit{Vidyo1} 16: \textit{Vidyo3} 17: \textit{Vidyo4} }
    \end{tabular}%
  \label{tab:time}%
  \vspace{-2.5em}
\end{table*}%

Next, we focus on modeling $\Delta\text{MSE}_n(k_n)$ for HEVC P frames. Here, Mean Absolute Deviation (MAD) is used to predict the potential quality enhancement of $\Delta\text{MSE}_n(k_n)$ at each CTU\footnote{It consumes 0.492 ms to calculate MADs for all CTUs in a 1080p frame using an Intel Core(TM) i7-4790K CPU. Thus, it takes much less time than QE-CNN.} because it is strongly correlated with $\Delta\text{MSE}_n(k_n)$, as discussed in the following. We define $m_n$ as the MAD value of the $n$-th CTU, and $S(m_n)$ is the rank of the descending sort of $\{m_n\}_{n=1}^N$ within a frame. Fig. \ref{train_P} shows the relationship between $S(m_n)$ and $\Delta\text{MSE}_n(k_n)$ over all training sequences. Here, we apply the normalized index $S(m_n)/N$ as the horizontal axis in Fig. \ref{train_P}, such that the relationship between $\Delta\text{MSE}_n(k_n)$ and $S(m_n)/N$ is suitable for different resolutions (with different numbers of CTUs). From Fig. \ref{train_P}, two facts can be observed: 1) $\Delta\text{MSE}_n(k_n=1)$ and $\Delta\text{MSE}_n(k_n=2)$ both decrease along with decreasing $m_n$, and 2) $\Delta\text{MSE}_n(k_n=2)$ decreases faster than $\Delta\text{MSE}_n(k_n=1)$ with decreasing $m_n$. Accordingly, we need to set a larger $k_n$ for CTUs with larger $m_n$. That is, the values of $k_n$ can be set as
\begin{equation}
k_n=\left\{
\begin{aligned}
2&,&\quad S(m_n) \leq N_2,  \\
1&,&\quad N_2 < S(m_n) \leq N_1+N_2,\\
0&,&\quad S(m_n) > N_1+N_2,
\end{aligned}
\right. \label{kn}
\end{equation}
where $N_1$ and $N_2$ represent the numbers of CTUs with $k_n=1$ and $k_n=2$, respectively. Given \eqref{kn}, formulation \eqref{formu} can be rewritten as follows:
\begin{eqnarray}
\label{formu2}
&&\max_{N_1, N_2} \sum_{S(m_n)=1}^{N_2}\!\!\!\!\!\! \Delta\text{MSE}_n(k_n=2) +\!\!\!\!\!\! \sum_{S(m_n)=N_2+1}^{N_2+N_1}\!\!\!\!\!\! \Delta\text{MSE}_n(k_n=1)\nonumber \\
&&\text{s.t.} \quad N_1\cdot t_n(k_n=1)+ N_2\cdot t_n(k_n=2)\leq T.
\end{eqnarray}
Then, to obtain the function of $\Delta\text{MSE}_n(k_n)$ in \eqref{formu2}, we utilize least-squares fitting of the second-order polynomial regression, and the fitting function is obtained as follows:
\begin{eqnarray}
\label{train_p}
\Delta\text{MSE}_n(k_n=1) = a_1\big(\frac{S(m_n)}{N}\big)^2-b_1\frac{S(m_n)}{N}+c_1, \label{1}\\
\Delta\text{MSE}_n(k_n=2) = a_2\big(\frac{S(m_n)}{N}\big)^2-b_2\frac{S(m_n)}{N}+c_2, \label{2}
\end{eqnarray}
where $a_1$, $a_2$, $b_1$, $b_2$, $c_1$ and $c_2$ are the coefficients presented in Table \ref{coe}. Table \ref{coe} also reports that all the R-square values of the fitting are above 0.99, verifying the effectiveness of fitting functions \eqref{1} and \eqref{2}.

Given \eqref{1} and \eqref{2}, the problem of \eqref{formu2} is a non-linear integer optimization problem. Therefore, we apply the branch-and-bound algorithm \cite{li2006nonlinear} to solve \eqref{formu2}. To avoid the overhead computational time consumed by solving formulation \eqref{formu2}, we establish a look-up table for the solutions to \eqref{formu2} at each specific $T$. Then, given $T$, we can simply obtain $N_1$ and $N_2$ by table look-up. Table \ref{tab:solution} presents the solution for 1080p sequences as an example of the table look-up. In Table \ref{tab:solution}, we use $T/T_{\max}$ instead of $T$ as the target. Note that $T_{\max} = N\cdot t_n(k_n=2)$ is the maximum time of our TQEO scheme, which applies QE-CNN to all CTUs. Additionally, $T_{\max}$ is a constant for each specific video. That is, the constraint in \eqref{formu2} can be represented by
\begin{eqnarray}
\label{formu3}
\quad \frac{N_1}{N}\cdot \frac{t_n(k_n=1)}{t_n(k_n=2)}+ \frac{N_2}{N}\leq \frac{T}{T_{\max}}
\end{eqnarray}
for establishing the look-up table. The reason is that the values of $t_n(k_n=1)$ and $t_n(k_n=2)$ may vary among different devices, but $t_n(k_n=1)/t_n(k_n=2) \approx 0.394 $ is almost unchanged. Therefore, $T/T_{\max}$ in Table \ref{tab:solution} makes the solution more general for different devices\footnote{In our experiments, we tested $t_n(k_n=1)/t_n(k_n=2)$ on three different devices, which are equipped with a GeForce GTX TITAN GPU, GeForce GTX 1080 GPU and GeForce GTX 1080 Ti GPU, respectively. We found that the values of $t_n(k_n=1)/t_n(k_n=2)$ are all within the range of 0.393 $\sim$ 0.395. Moreover, in practice, $T_{\max}$ can be simply estimated by recording the computational time of QE-CNN-P on a few CTUs.}.

Finally, the solution to \eqref{formu} can be obtained for the P frames of HEVC compressed videos. As a result, the computational time for quality enhancement can be controlled for both I and P frames with maximal enhancement on video quality.

\begin{figure*}[!t]
\centering
\vspace{-2.5em}
\subfigure{\includegraphics[width = 0.24\linewidth]{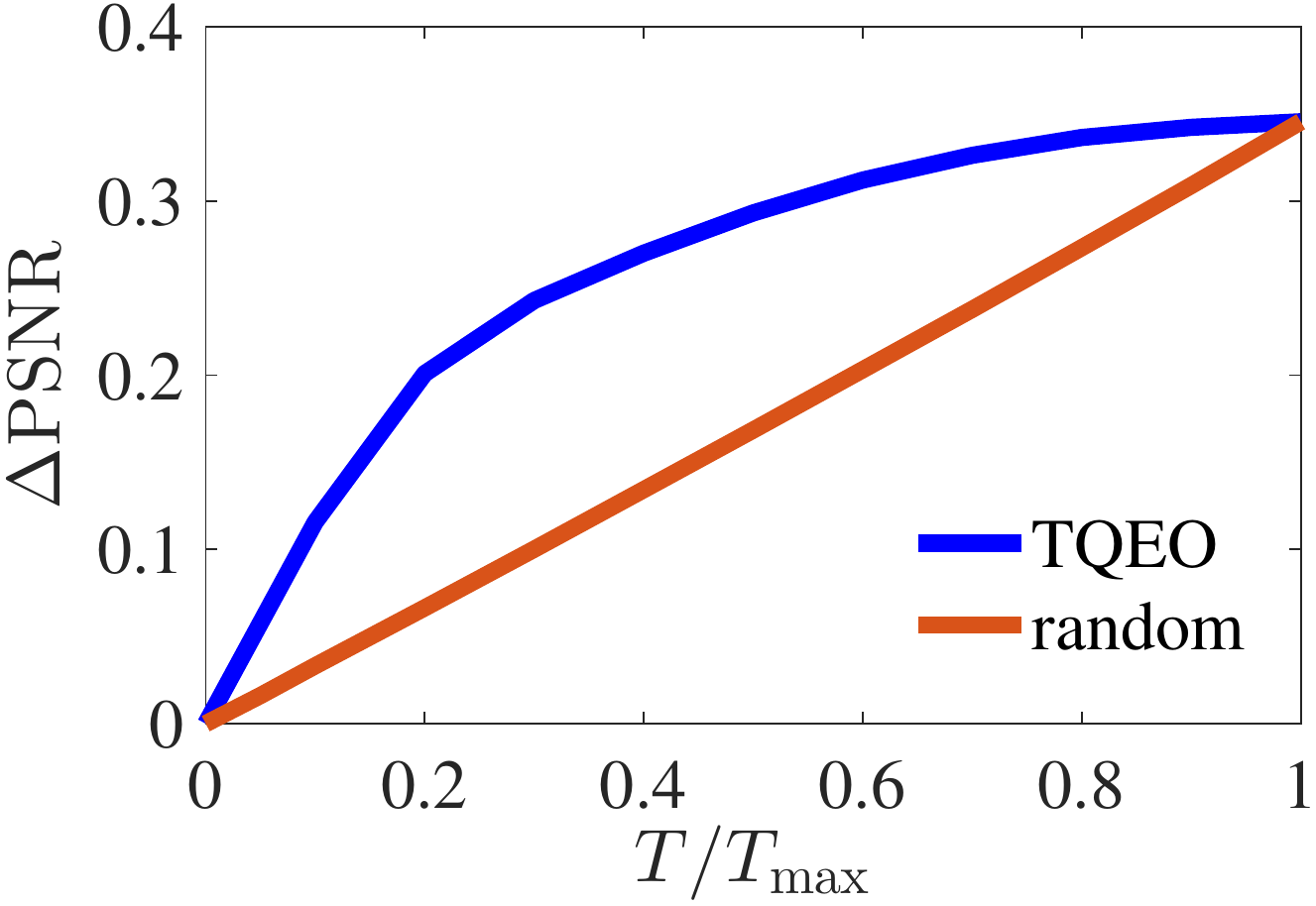}}
\subfigure{\includegraphics[width = 0.24\linewidth]{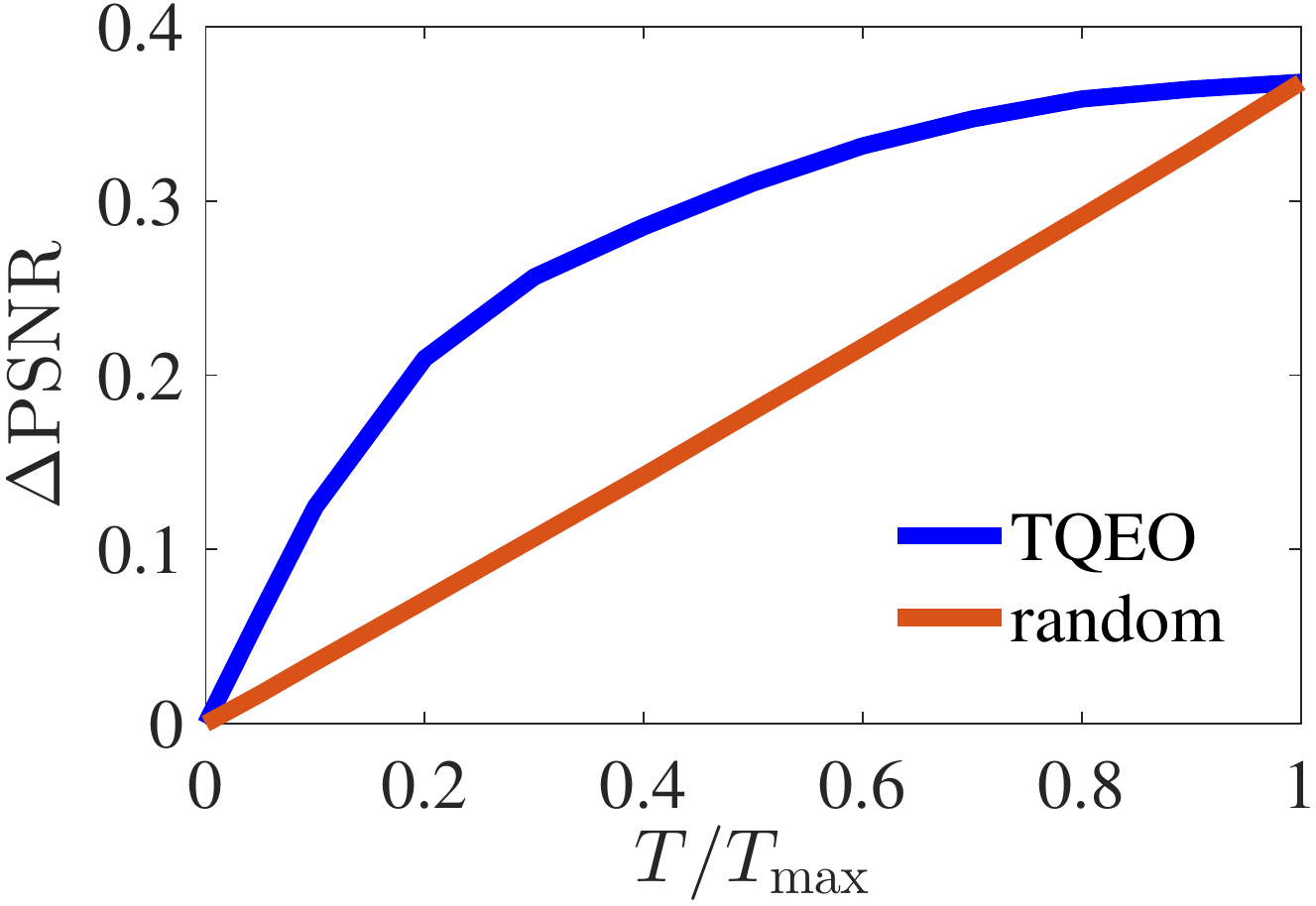}}
\subfigure{\includegraphics[width = 0.24\linewidth]{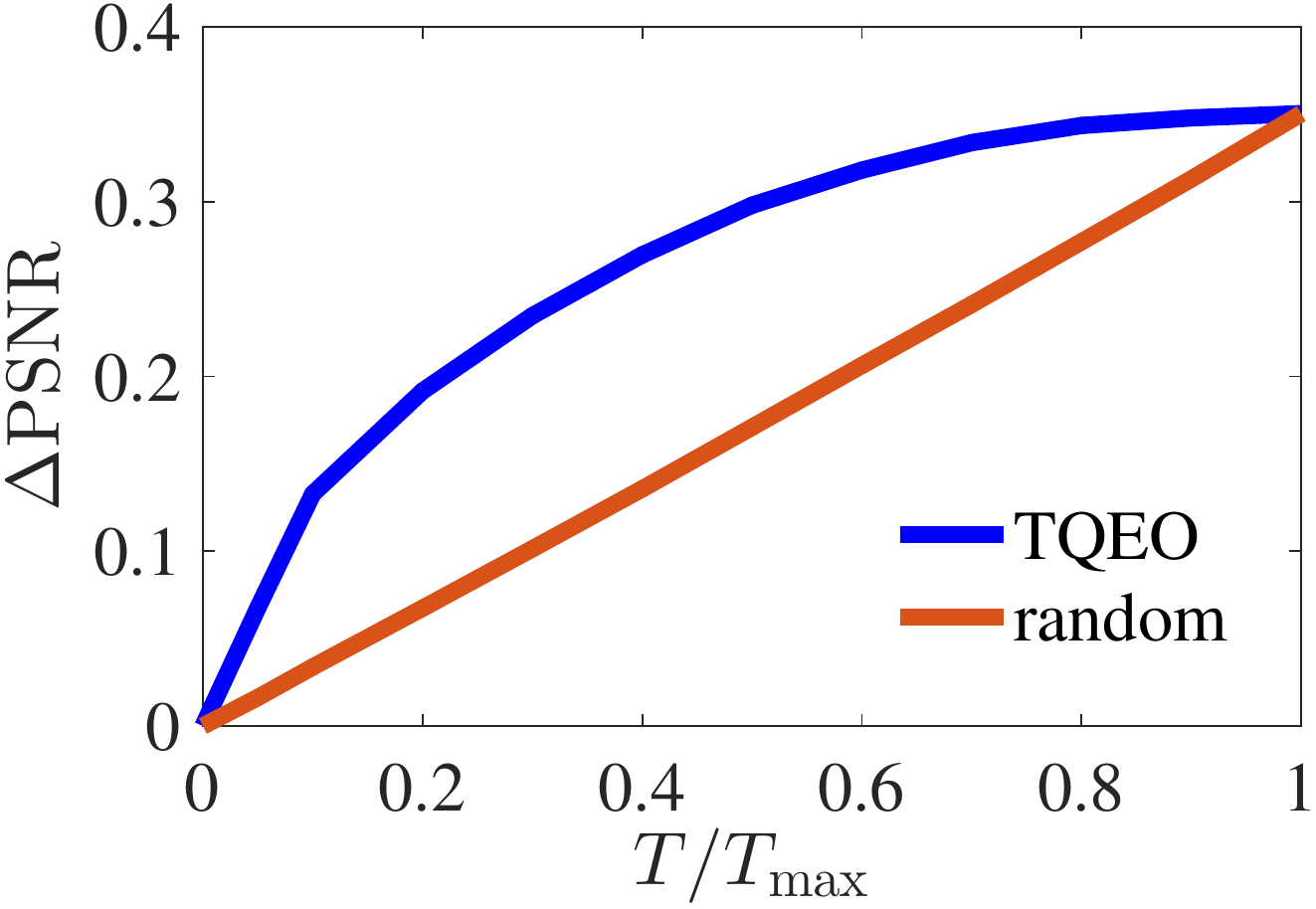}}
\subfigure{\includegraphics[width = 0.24\linewidth]{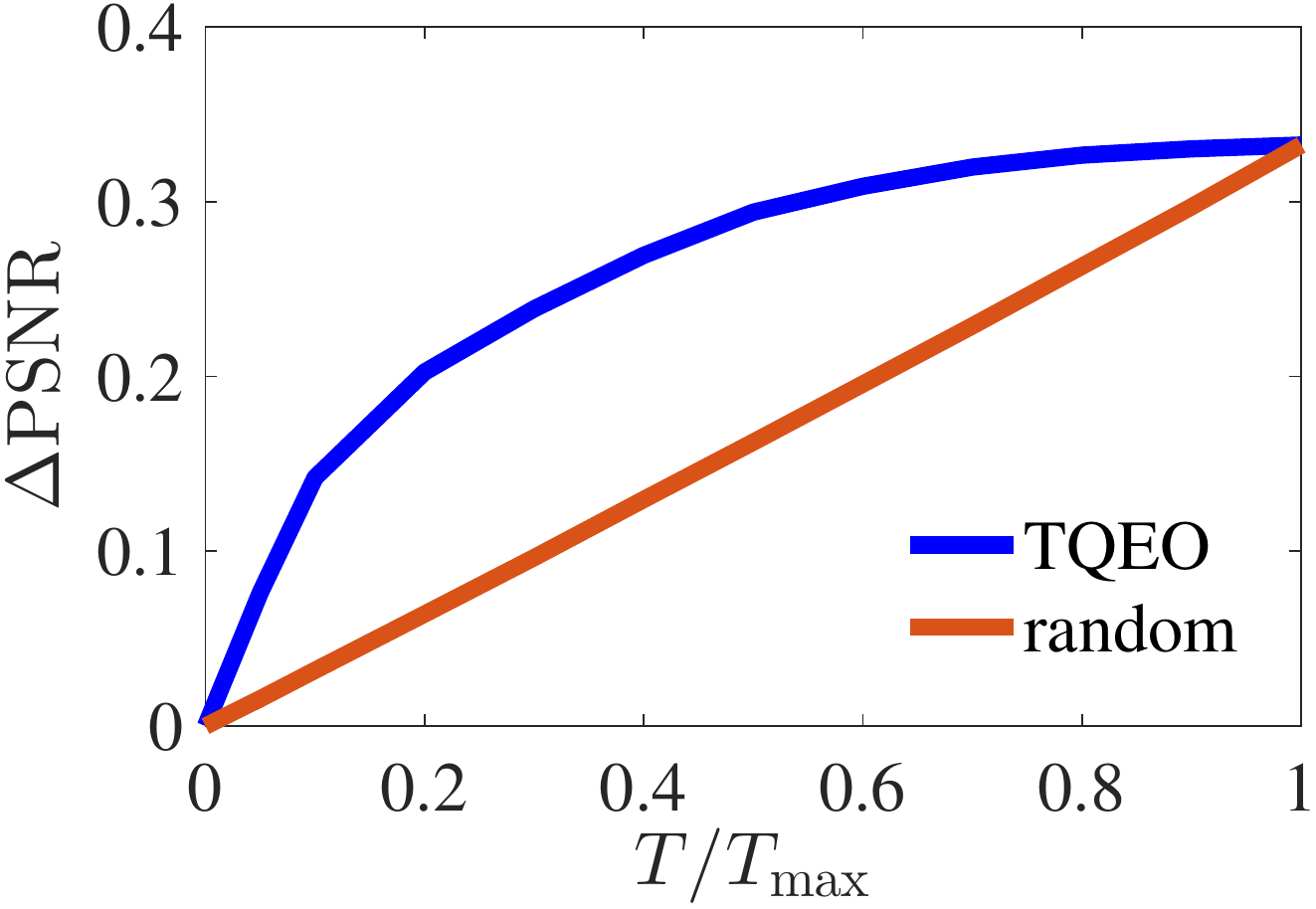}}
\vspace{-1.2em}
\caption{Time-constrained quality enhancement performance averaged among all the test sequences.}\label{ave_QEO}
\vspace{-.7em}
\end{figure*}

\begin{figure*}[!t]
\centering
\subfigure{\includegraphics[width = 0.24\linewidth]{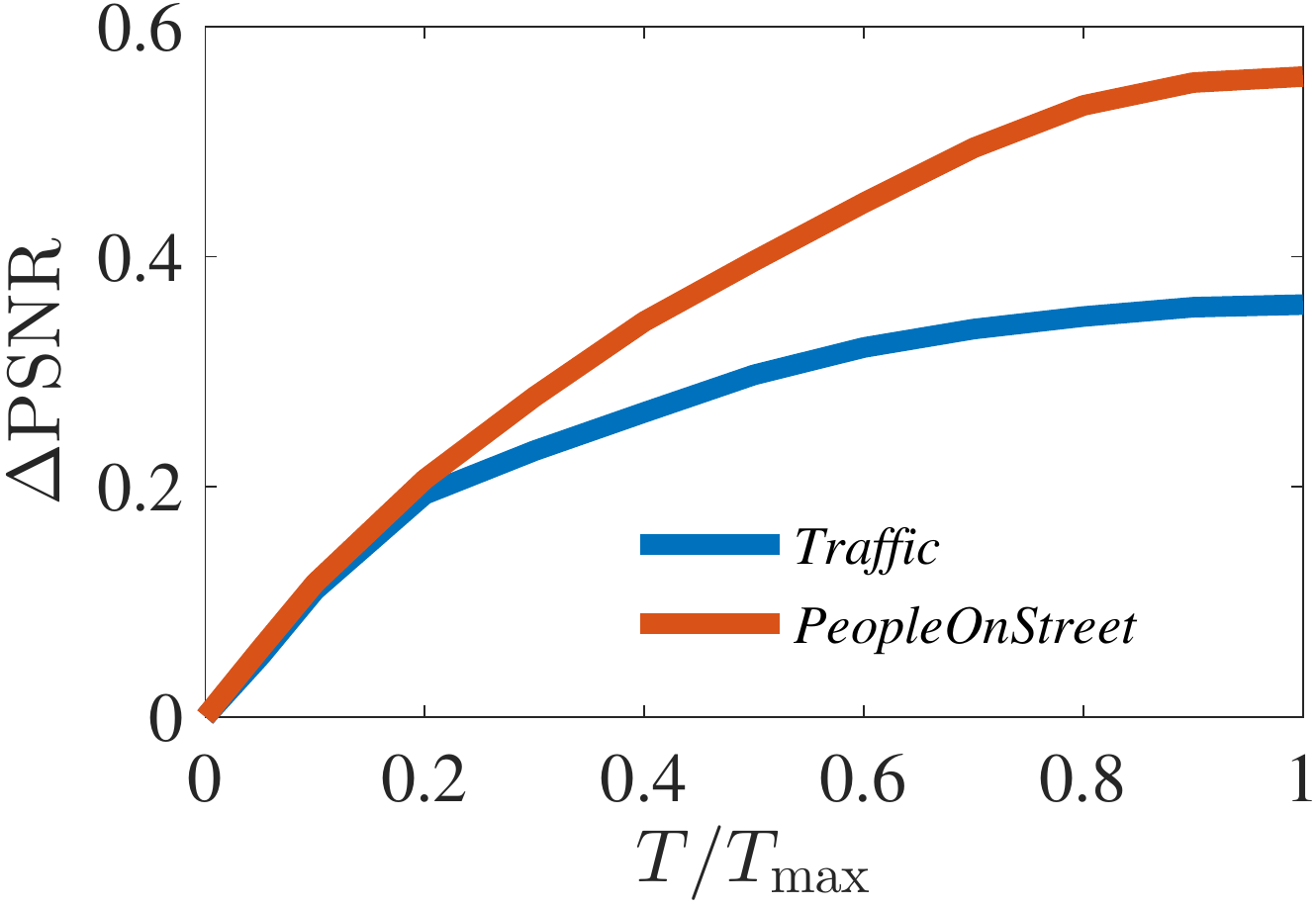}}
\subfigure{\includegraphics[width = 0.24\linewidth]{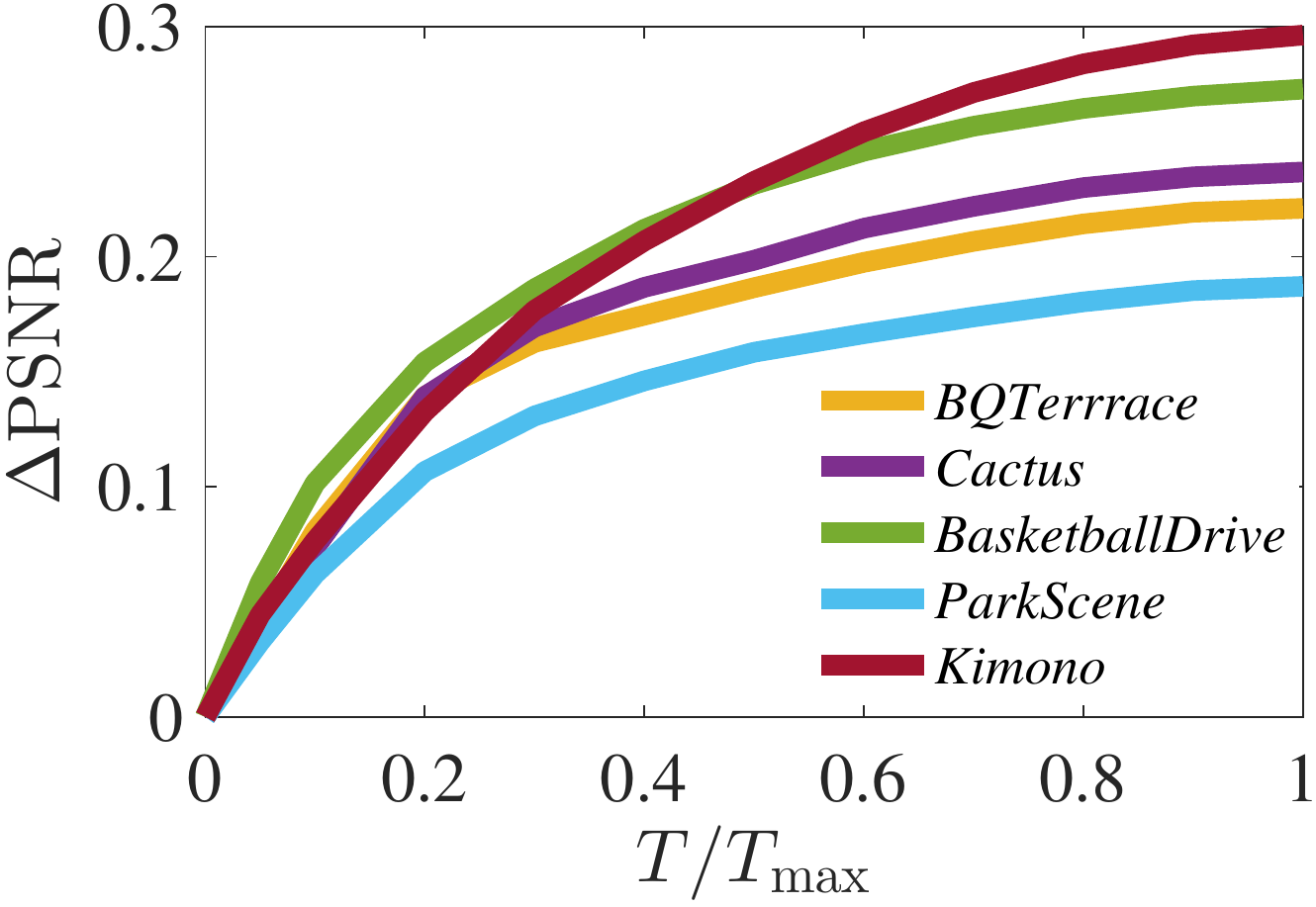}}
\subfigure{\includegraphics[width = 0.24\linewidth]{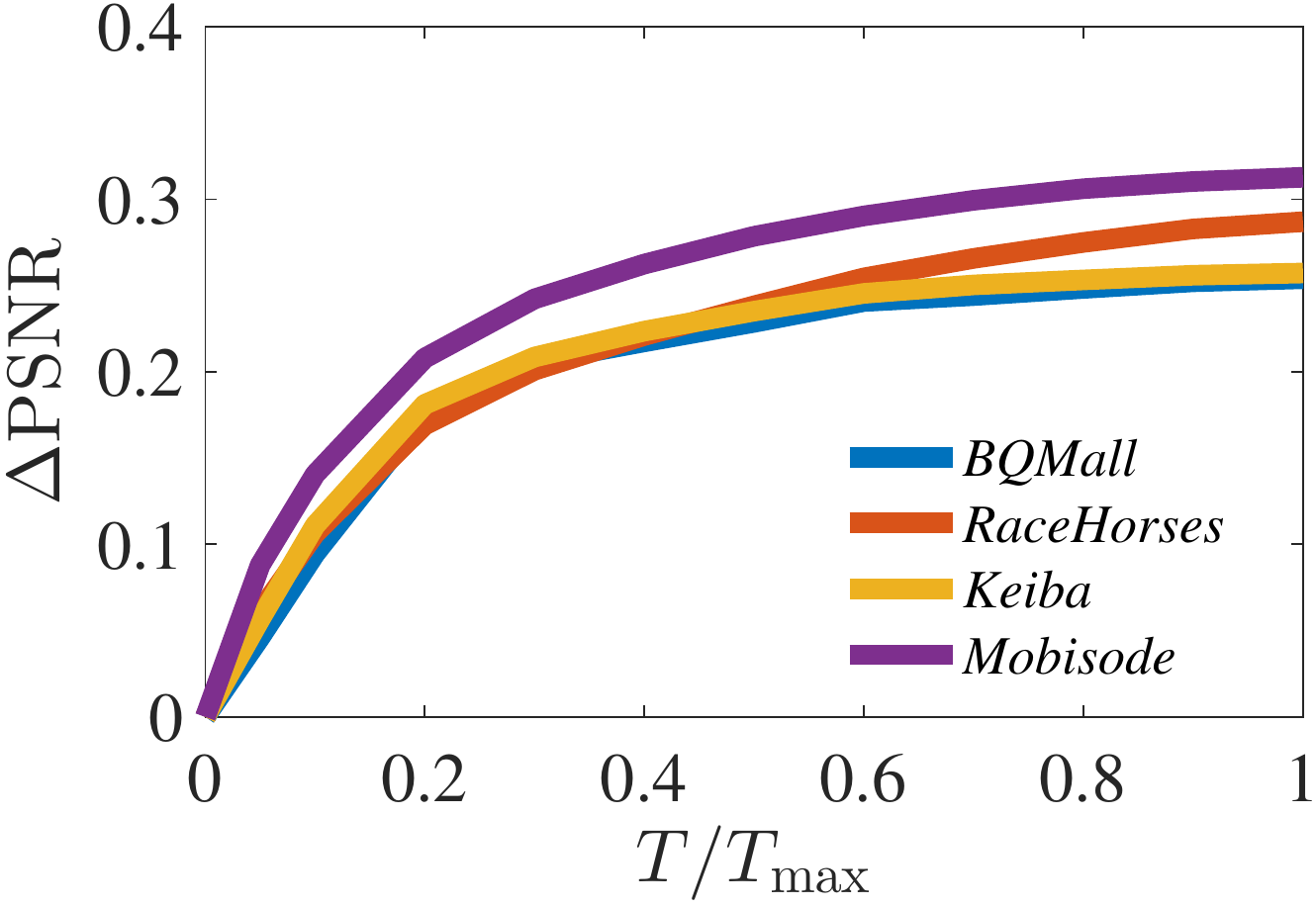}}
\subfigure{\includegraphics[width = 0.24\linewidth]{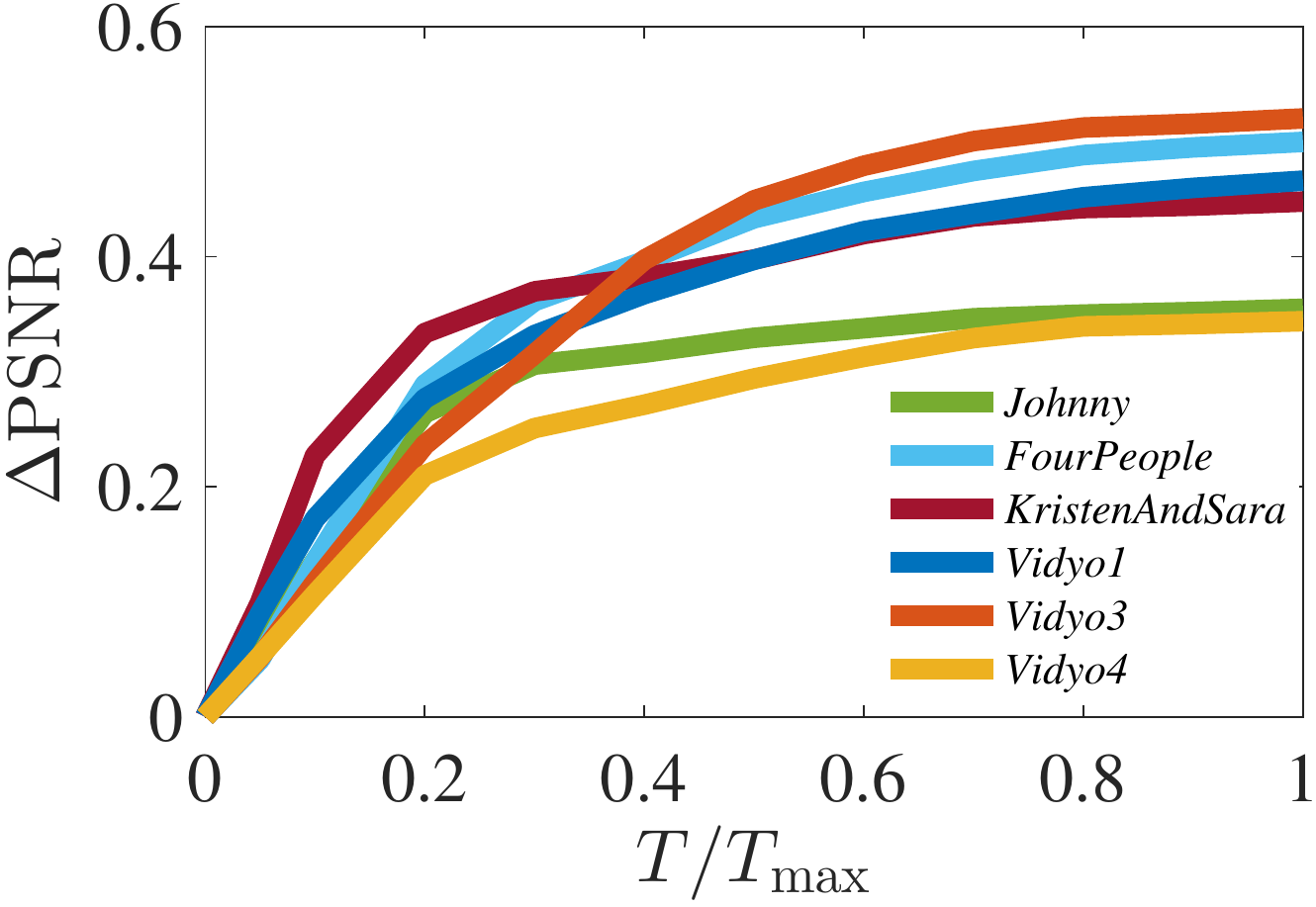}}
\vspace{-1.2em}
\caption{Time-constrained quality enhancement performance on each test sequence at QP = 32.}\label{QEO}
\vspace{-1.7em}
\end{figure*}

\vspace{-1em}
\section{Experiments for our TQEO scheme}\label{EX-TC}

In this section, we validate the performance of our TQEO scheme from the aspects of both control accuracy and quality enhancement evaluated under different time constraints. The experiments are conducted on an Ubuntu PC with one GeForce GTX 1080 GPU. The HEVC test sequences are the same as in Section \ref{EX-QE}.
Note that, there is some decoding information required for QE-CNN and TQEO. Specifically, in our QE-CNN method, the QP value of each frame is required to select the trained model used for this frame. In TQEO scheme,the bit allocation ($b_n$) of each CTU at the I frames is required to sort the priority in enhancing quality under time constraint. No other decoding information is required.

\vspace{-1em}
\subsection{Accuracy of computational time control}

We now evaluate the accuracy of computational time control of our TQEO scheme to validate its effectiveness in controlling the time to a constraint. Since no other work focuses on time control of quality enhancement, we do not compare the control accuracy with other methods. Table \ref{tab:time} presents the time control accuracy of our TQEO scheme. In Table \ref{tab:time}, we present the results of actual time divided by the maximum time $T_{\max}$. As shown in Table \ref{tab:time}, the control error is up to $1.89\%$ (\textit{RaceHorses}, QP = 42, $20\%$), and most of the errors are less than $0.200\%$. Table \ref{tab:time} also tabulates the Mean Absolute Error (MAE), which is calculated as
\begin{eqnarray}
\text{MAE} = \frac{|\Delta T|}{T_{\max}},
\end{eqnarray}
averaged over all the test sequences. Here, $\Delta T$ indicates the error between the target and actual time. It is apparent that the average MAEs of TQEO in most cases are below $0.150\%$, and the highest value of MAE is only $0.333\%$. In conclusion, our TQEO scheme performs well in controlling the computational time of quality enhancement under a time constraint.

\vspace{-1em}
\subsection{Quality enhancement under different time constraints}

Next, we focus on assessing the quality enhancement of our TQEO scheme under different time constraints. As in Section \ref{EX-QE}, quality enhancement is measured in terms of Y-PSNR improvement ($\Delta$PSNR). We compare the $\Delta$PSNR of our TQEO scheme with a baseline that applies the QE-CNN method on randomly selected CTUs. Fig. \ref{ave_QEO} shows the $\Delta$PSNR averaged among all 17 test sequences, along with different time constraints. As shown, under each time constraint, our TQEO scheme achieves a higher $\Delta$PSNR than the baseline. Specifically, the average $\Delta$PSNR of our TQEO scheme reaches $58.2\%$ of the maximum $\Delta$PSNR at QP = 32, with only $20\%$ of the computational time of $T_{\max}$. When the time constraint is half of $T_{\max}$, our TQEO scheme is able to improve the Y-PSNR by 0.2931 dB on average, which is $84.9\%$ of the maximum $\Delta$PSNR. Moreover, Fig. \ref{QEO} presents the results on all test sequences at QP = 32. Similar to the average $\Delta$PSNR curves, we can observe that the curves are up-convex for quality enhancement under different time constraints. Consequently, the quality enhancement of our TQEO scheme significantly outperforms the baseline of enhancing randomly selected CTUs, whose $\Delta$PSNR is almost linearly related to $T$. That is, our TQEO scheme succeeds in optimizing the quality enhancement given the limited computational time.

Furthermore, we also investigate whether our TQEO scheme leads to different sharpness in different parts of frames. We conduct a subjective experiment for AR-CNN \cite{dong2015compression}, DCAD \cite{Wang2017A}, QE-CNN and QE-CNN with 50\% computational complexity. In this experiment, for each sequence, the subjects are asked whether different parts of the frames are differently sharp. The score of 1 indicates that the viewer observe different sharpness in different parts of frames, and 0 stands for no different sharpness observed in the video. There are 12 non-expert subjects participating in this experiment. Table \ref{tab:sub} shows the average scores of sharpness difference of the 12 subjects. It can be seen in Table VII that, the scores of the methods on the whole frames (AR-CNN \cite{dong2015compression}, DCAD \cite{Wang2017A}, QE-CNN) and the partly post-processing method (QE-CNN with 50\% complexity) are rather similar. This indicates that our TQEO scheme does not lead to additional sharpness difference. It is probably because the HEVC compression has already led to quality difference at different parts of frames, which is much larger than that caused by quality enhancement.

Besides, the DMOS experiments are also conducted for TQEO. The DMOS value in Table \ref{tab:sub} shows that QE-CNN with 50\% complexity outperforms AR-CNN \cite{dong2015compression} and DCAD \cite{Wang2017A} for most sequences. The average DMOS value of QE-CNN with 50\% complexity is 49.22, which is smaller than those AR-CNN (54.42) and DCAD (51.86), indicating the better subjective quality. These results validate that when frames are partly enhanced using our TQEO scheme, the subjective quality does not significantly decrease and is still better than the state-of-the-art methods \cite{dong2015compression,Wang2017A}.

\vspace{-1em}
\subsection{Real-time Implementation}

\begin{table*}[!t]
\vspace{-3.5em}
  \centering
  \scriptsize
  \caption{Performance of our prototype.}
\vspace{-1em}
\begin{tabular}{|c|c||c|c|c|c||c|c|c|c||c|}
\hline
\multirow{2}[4]{*}{Class} &
  \multirow{2}[4]{*}{Sequence} &
  \multicolumn{4}{c||}{Actual time (s)} &
  \multicolumn{4}{c||}{$\Delta$PSNR (dB)} &
  \multirow{2}[4]{*}{BD-rate (\%)}
  \\
\cline{3-10} &
   &
  QP = 32 &
  QP = 37 &
  QP = 42 &
  QP = 47 &
  QP = 32 &
  QP = 37 &
  QP = 42 &
  QP = 47 &

  \\
\hline
\multirow{4}[2]{*}{C} &
  \textit{RaceHorses} &
  9.81 &
  9.81 &
  9.81 &
  9.83 &
  0.2169 &
  0.2262 &
  0.2121 &
  0.2117 &
  -6.8282
  \\
\cline{2-11} &
  \textit{Keiba} &
  9.93 &
  9.93 &
  9.87 &
  9.86 &
  0.2211 &
  0.2320 &
  0.2290 &
  0.3010 &
  -6.1855
  \\
\cline{2-11} &
  \textit{Mobisode} &
  9.93 &
  9.91 &
  9.97 &
  9.71 &
  0.2580 &
  0.2714 &
  0.2274 &
  0.2248 &
  -5.9942
  \\
\cline{2-11} &
  \textit{BQMall} &
  9.98 &
  9.97 &
  9.94 &
  10.0  &
  0.1654 &
  0.1774 &
  0.1427 &
  0.1493 &
  -3.8446
  \\
\hline
\multirow{6}[2]{*}{E/E'} &
  \textit{Johnny} &
  9.90 &
  10.0  &
  9.91 &
  9.90 &
  0.1026 &
  0.1154 &
  0.1107 &
  0.1992 &
  -2.8413
  \\
\cline{2-11} &
  \textit{FourPeople} &
  9.91 &
  9.97 &
  9.90 &
  9.99 &
  0.0956 &
  0.1119 &
  0.1392 &
  0.1011 &
  -2.4498
  \\
\cline{2-11} &
  \textit{KristenAndSara} &
  9.88 &
  9.99 &
  9.91 &
  10.0  &
  0.1754 &
  0.2284 &
  0.2531 &
  0.2617 &
  -4.8185
  \\
\cline{2-11} &
  \textit{Vidyo1} &
  9.97 &
  10.0  &
  9.99 &
  10.0  &
   0.1356 &
  0.1409 &
  0.1533 &
  0.1622 &
  -3.0907
  \\
\cline{2-11} &
  \textit{Vidyo3} &
  10.0  &
  9.98 &
  9.99 &
  10.0  &
  0.0887 &
  0.0674 &
  0.0565 &
  0.0729 &
  -1.4138
  \\
\cline{2-11} &
  \textit{Vidyo4} &
  9.92 &
  9.88 &
  9.88 &
  10.0  &
  0.0823 &
  0.0873 &
  0.1021 &
  0.1231 &
  -2.2120
  \\
\hline
\end{tabular}%
\vspace{-2em}
  \label{tab:proto}%
\end{table*}%

Finally, we design a prototype to achieve real-time quality enhancement of HEVC compressed videos by using our TQEO scheme. Our prototype is implemented on an Ubuntu PC with four GeForce GTX 1080 GPUs. In our prototype, the computational time target $T$ in formulation \eqref{formu} needs to satisfy the real-time constraint\footnote{Note that we did not take the decoding time into consideration in our prototype, because the latest HEVC decoder achieves very fast decoding. For example, the decoder proposed in \cite{zhou201614} is able to decode HEVC LDP videos at a speed of 4 Gpixels/s (0.14 s for a 10 s 720p@60 Hz sequence).}, e.g., $T = 16.67$ ms per each frame for 60 Hz sequences.

\begin{figure}[!t]
\centering
\subfigure{\includegraphics[width = 0.49\linewidth]{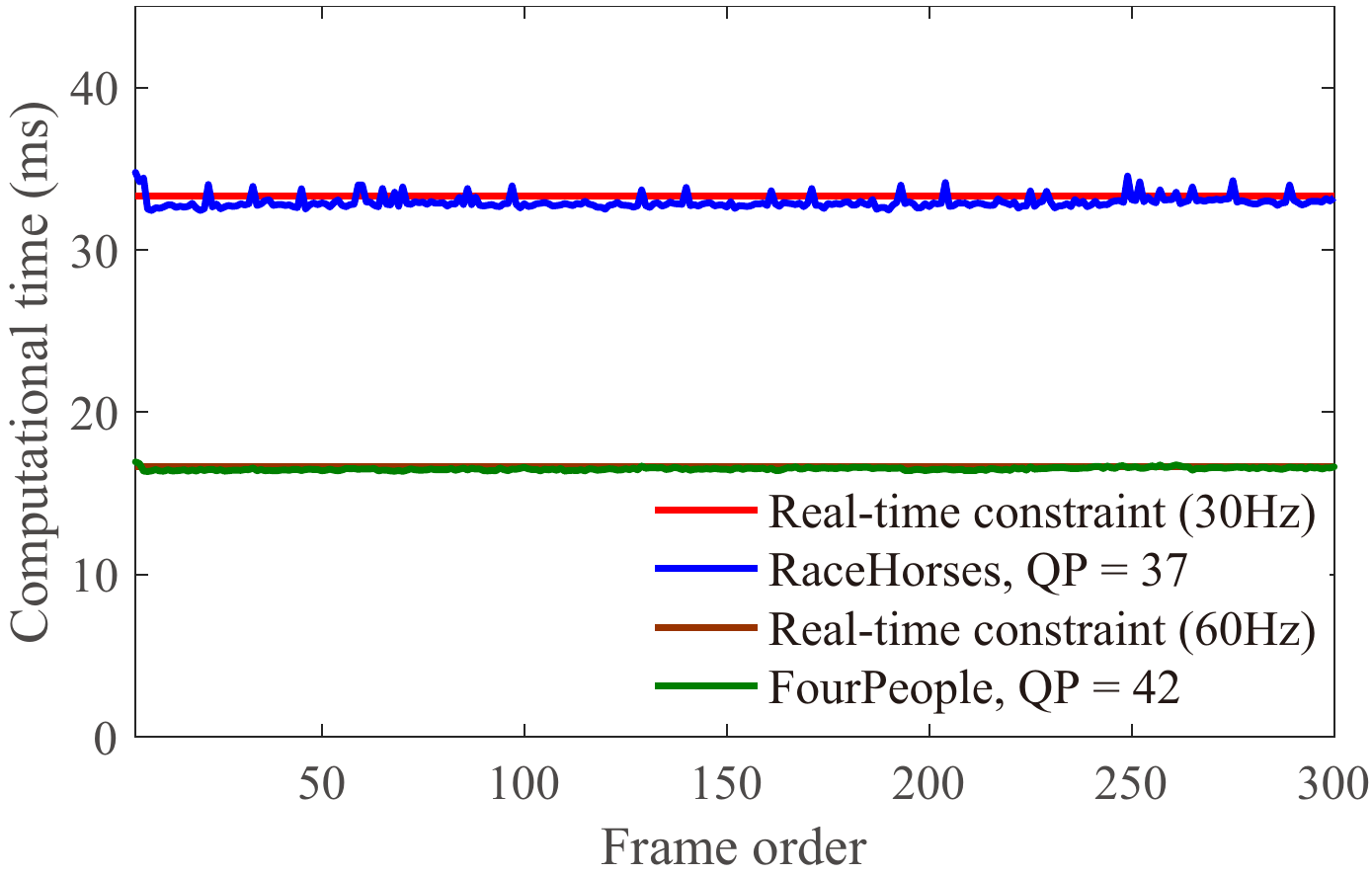}}
\subfigure{\includegraphics[width = 0.49\linewidth]{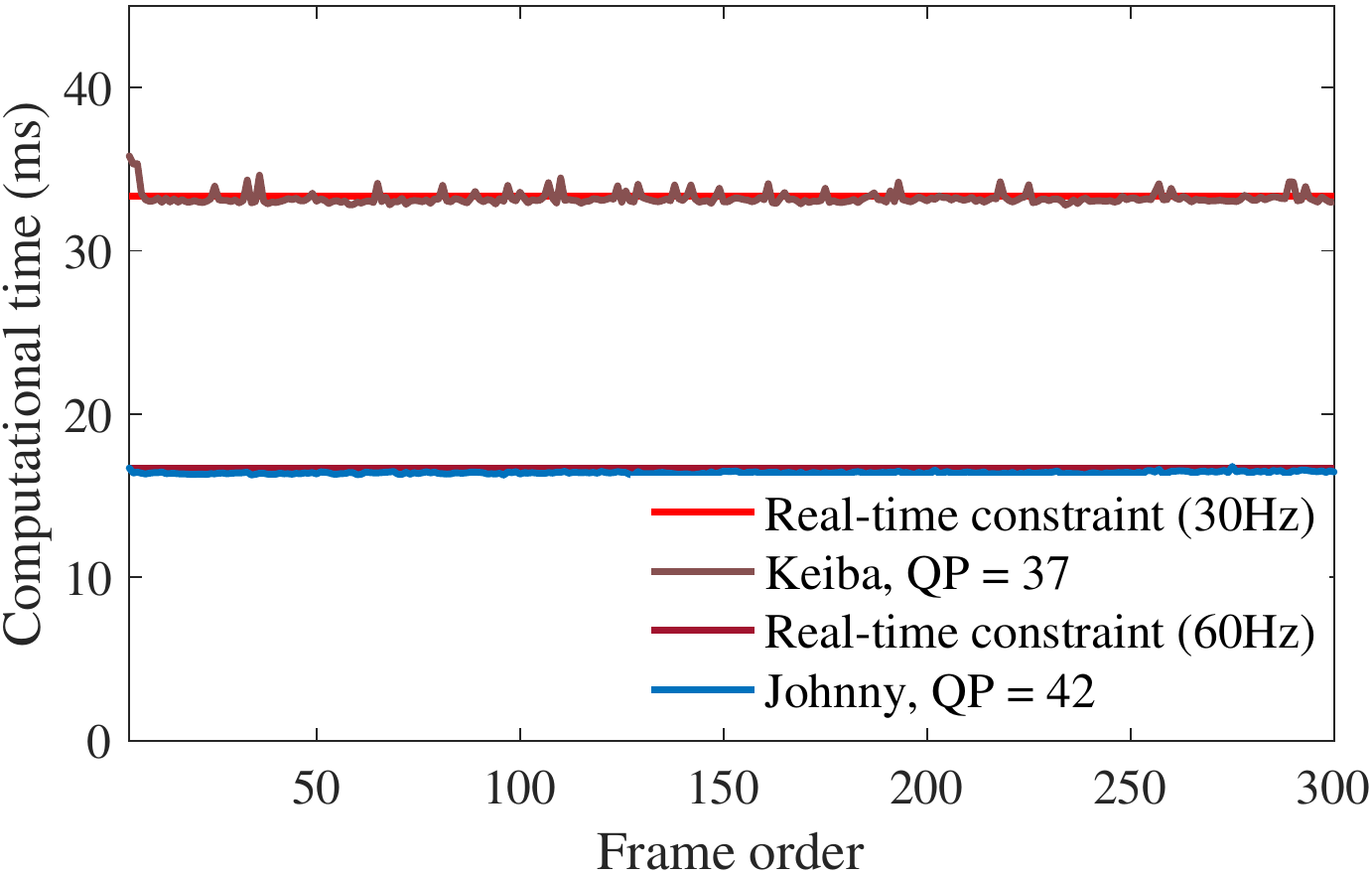}}
\vspace{-2em}
\caption{Frame-level computational time of our prototype.}\label{time_frame}
\vspace{-1.5em}
\end{figure}

We evaluate the performance our prototype on 10 test sequences. The duration of each test sequence is 10 s. Table \ref{tab:proto} presents the actual computational time and quality enhancement for our prototype. We can observe that the actual time used for quality enhancement satisfies the real-time constraint. Moreover, the control error of the computational time is very small, which is up to 0.19 s, far less than the video duration of 10 s. The average control error is only 0.073 s, i.e., 0.73\% of the 10 s duration. Fig. \ref{time_frame} shows the computational time of our prototype at each frame of sequences \textit{RaceHorses}, \textit{FourPeople}, \textit{Keiba} and \textit{Johnny}. As shown in Fig. \ref{time_frame}, the frame-level control error of our prototype is also small, indicating high fluency when displaying the enhanced HEVC videos.

The quality enhancement performance is shown in Table \ref{tab:proto}. As shown, at QP = 47, our prototype has a 0.2458 dB $\Delta$PSNR on 480p@30 Hz sequences on average (\textit{RaceHorses}, \textit{Keiba} and \textit{Mobisode}), and it achieves 0.1493 dB on the 480p@60 Hz sequence (\textit{BQMall}). For 720p@60 Hz (Class E/E') sequences, the $\Delta$PSNR is 0.1534 dB on average at QP = 47. Similar results can be found for the other QPs. Furthermore, our prototype can achieve BD-rate savings of up to 6.83\%, and an average of 6.34\% for 480p@30 Hz sequences, in our real-time prototype. The BD-rate savings for 720p@60 Hz (Class E/E') sequences is 2.80\% on average when achieving real-time enhancement. To summarize, our prototype is effective in both time control accuracy and quality enhancement for HEVC compressed videos in the real-time implementation.

\vspace{-1em}
\section{Conclusion}

This paper has proposed the QE-CNN method at the decoder side, to improve the quality of HEVC videos. Our QE-CNN method learns to reduce the artifacts of both intra- and inter-coding, rather than only reducing intra-mode compression artifacts in the existing CNN-based quality enhancement methods. Specifically, two networks, i.e., QE-CNN-I and QE-CNN-P, were proposed to learn the features of intra- and inter-mode distortion, respectively. As such, our QE-CNN method can enhance the quality for both I and P/B frames of HEVC compressed videos. The experimental results have shown that the proposed QE-CNN method is able to advance the state-of-the-art quality enhancement of HEVC videos.

However, the QE-CNN method introduces heavy computational complexity. To make the quality enhancement adaptive to different time constraints, we further proposed the TQEO scheme for controlling the computational time of our QE-CNN method. We first established the TQEO formulation, which maximizes the enhanced quality under a time constraint by selecting some of the CTUs for quality enhancement. Then, two solutions were derived for the established formulation for the intra- and inter-coding frames of HEVC compressed videos. The experimental results verified the effectiveness of the proposed TQEO scheme in terms of control accuracy and quality enhancement. Finally, a prototype was established to implement our TQEO scheme in a real-time application of quality enhancement.

In our prototype, at the current stage, four high-end GPUs are adopted to achieve real-time enhancement. This prototype shows an example that our TQEO scheme is able to control the computational time of quality enhancement to a specific target, adjusting to the computational resources. Fortunately, there are plenty of acceleration hardwares to speed up the CNN networks. For example, the Cambricon-X accelerator has more than 7 times speed of GPU with much lower energy \cite{zhang2016cambricon}. Recently, the Tensor Processing Unit (TPU) \cite{jouppi2017datacenter} developed by Google achieves 15 to 30 times speed compared with GPU, with smaller size and lower power consumption. Applying our QE-CNN method and TQEO scheme on these Application Specific Integrated Circuits (ASIC) can be seen as an interesting further work.

\ifCLASSOPTIONcaptionsoff
  \newpage
\fi

%
%

\bibliographystyle{IEEEtran}
\bibliography{icme2016template}

\begin{thebibliography}{10}
\providecommand{\url}[1]{#1}
\csname url@samestyle\endcsname
\providecommand{\newblock}{\relax}
\providecommand{\bibinfo}[2]{#2}
\providecommand{\BIBentrySTDinterwordspacing}{\spaceskip=0pt\relax}
\providecommand{\BIBentryALTinterwordstretchfactor}{4}
\providecommand{\BIBentryALTinterwordspacing}{\spaceskip=\fontdimen2\font plus
\BIBentryALTinterwordstretchfactor\fontdimen3\font minus
  \fontdimen4\font\relax}
\providecommand{\BIBforeignlanguage}[2]{{%
\expandafter\ifx\csname l@#1\endcsname\relax
\typeout{** WARNING: IEEEtran.bst: No hyphenation pattern has been}%
\typeout{** loaded for the language `#1'. Using the pattern for}%
\typeout{** the default language instead.}%
\else
\language=\csname l@#1\endcsname
\fi
#2}}
\providecommand{\BIBdecl}{\relax}
\BIBdecl

\bibitem{sullivan2012overview}
G.~J. Sullivan, J.-R. Ohm, W.-J. Han, and T.~Wiegand, ``Overview of the high
  efficiency video coding ({HEVC}) standard,'' \emph{IEEE TCSVT}, pp.
  1649--1668, 2012.

\bibitem{Tan2016}
T.~K. Tan, R.~Weerakkody, M.~Mrak, N.~Ramzan, V.~Baroncini, J.~R. Ohm, and
  G.~J. Sullivan, ``Video quality evaluation methodology and verification
  testing of hevc compression performance,'' \emph{IEEE TCSVT}, pp. 76--90, Jan
  2016.

\bibitem{liew2004blocking}
A.-C. Liew and H.~Yan, ``Blocking artifacts suppression in block-coded images
  using overcomplete wavelet representation,'' \emph{IEEE TCSVT}, pp. 450--461,
  2004.

\bibitem{foi2007pointwise}
A.~Foi, V.~Katkovnik, and K.~Egiazarian, ``Pointwise shape-adaptive {DCT} for
  high-quality denoising and deblocking of grayscale and color images,''
  \emph{IEEE TIP}, 2007.

\bibitem{jancsary2012loss}
J.~Jancsary, S.~Nowozin, and C.~Rother, ``Loss-specific training of
  non-parametric image restoration models: A new state of the art,'' in
  \emph{ECCV}, 2012.

\bibitem{chang2014reducing}
H.~Chang, M.~K. Ng, and T.~Zeng, ``Reducing artifacts in {JPEG} decompression
  via a learned dictionary,'' \emph{IEEE TSP}, pp. 718--728, 2014.

\bibitem{jung2012image}
C.~Jung, L.~Jiao, H.~Qi, and T.~Sun, ``Image deblocking via sparse
  representation,'' \emph{Signal Processing: Image Communication}, pp.
  663--677, 2012.

\bibitem{wang2013adaptive}
C.~Wang, J.~Zhou, and S.~Liu, ``Adaptive non-local means filter for image
  deblocking,'' \emph{Signal Processing: Image Communication}, pp. 522--530,
  2013.

\bibitem{dong2015compression}
C.~Dong, Y.~Deng, C.~Change~Loy, and X.~Tang, ``Compression artifacts reduction
  by a deep convolutional network,'' in \emph{ICCV}, 2015.

\bibitem{wang2016d3}
Z.~Wang, D.~Liu, S.~Chang, Q.~Ling, and T.~S. Huang, ``D3: Deep dual-domain
  based fast restoration of {JPEG}-compressed images,'' \emph{arXiv preprint
  arXiv:1601.04149}, 2016.

\bibitem{lecun1998gradient}
Y.~LeCun, L.~Bottou, Y.~Bengio, and P.~Haffner, ``Gradient-based learning
  applied to document recognition,'' \emph{Proceedings of the IEEE}, pp.
  2278--2324, 1998.

\bibitem{Chen2017Trainable}
Y.~Chen and T.~Pock, ``Trainable nonlinear reaction diffusion: A flexible
  framework for fast and effective image restoration,'' \emph{IEEE TPAMI}, pp.
  1256--1272, 2017.

\bibitem{han2015high}
Q.~Han and W.-K. Cham, ``High performance loop filter for {HEVC},'' in
  \emph{ICIP}, 2015.

\bibitem{zhang2016structure}
J.~Zhang, C.~Jia, N.~Zhang, S.~Ma, and W.~Gao, ``Structure-driven adaptive
  non-local filter for high efficiency video coding ({HEVC}),'' in \emph{DCC},
  2016.

\bibitem{park2016cnn}
W.-S. Park and M.~Kim, ``{CNN}-based in-loop filtering for coding efficiency
  improvement,'' in \emph{IVMSP}, 2016.

\bibitem{dai2017convolutional}
Y.~Dai, D.~Liu, and F.~Wu, ``A convolutional neural network approach for
  post-processing in {HEVC} intra coding,'' in \emph{MMM}, 2017.

\bibitem{Wang2017A}
T.~Wang, M.~Chen, and H.~Chao, ``A novel deep learning-based method of
  improving coding efficiency from the decoder-end for hevc,'' in \emph{Data
  Compression Conference}, 2017.

\bibitem{Yang2017coding}
R.~Yang, M.~Xu, and Z.~Wang, ``Ddecoder-side {HEVC} quality enhancement with
  scalable convolutional neural network,'' in \emph{ICME}, 2017.

\bibitem{krizhevsky2012imagenet}
A.~Krizhevsky, I.~Sutskever, and G.~E. Hinton, ``Imagenet classification with
  deep convolutional neural networks,'' in \emph{NIPS}, 2012.

\bibitem{karpathy2014large}
A.~Karpathy, G.~Toderici, S.~Shetty, T.~Leung, R.~Sukthankar, and L.~Fei-Fei,
  ``Large-scale video classification with convolutional neural networks,'' in
  \emph{CVPR}, 2014.

\bibitem{girshick2014rich}
R.~Girshick, J.~Donahue, T.~Darrell, and J.~Malik, ``Rich feature hierarchies
  for accurate object detection and semantic segmentation,'' in \emph{CVPR},
  2014.

\bibitem{long2015fully}
J.~Long, E.~Shelhamer, and T.~Darrell, ``Fully convolutional networks for
  semantic segmentation,'' in \emph{CVPR}, 2015.

\bibitem{Dong2014learning}
C.~Dong, C.~L. Chen, K.~He, and X.~Tang, ``Learning a deep convolutional
  network for image super-resolution,'' in \emph{ECCV}, 2014.

\bibitem{li2016lagrangian}
S.~Li, C.~Zhu, Y.~Gao, Y.~Zhou, F.~Dufaux, and M.-T. Sun, ``Lagrangian
  multiplier adaptation for rate-distortion optimization with inter-frame
  dependency,'' \emph{IEEE TCSVT}, pp. 117--129, 2016.

\bibitem{gao2017temporal}
Y.~Gao, C.~Zhu, S.~Li, and T.~Yang, ``Temporal dependent rate-distortion
  optimization for low-delay hierarchical video coding,'' \emph{IEEE TIP},
  2017.

\bibitem{correa2011complexity}
G.~Corr{\^e}a, P.~Assuncao, L.~Agostini, and L.~A. da~Silva~Cruz, ``Complexity
  control of high efficiency video encoders for power-constrained devices,''
  \emph{IEEE TCE}, pp. 1866--1874, 2011.

\bibitem{deng2014complexity}
X.~Deng, M.~Xu, S.~Li, and Z.~Wang, ``Complexity control of hevc based on
  region-of-interest attention model,'' in \emph{VCIP}, 2014.

\bibitem{deng2016subjective}
X.~Deng, M.~Xu, L.~Jiang, X.~Sun, and Z.~Wang, ``Subjective-driven complexity
  control approach for {HEVC},'' \emph{IEEE TCSVT}, pp. 91--106, 2016.

\bibitem{langroodi2015decoder}
M.~J. Langroodi, J.~Peters, and S.~Shirmohammadi, ``Decoder-complexity-aware
  encoding of motion compensation for multiple heterogeneous receivers,''
  \emph{ACM TOMM}, p.~46, 2015.

\bibitem{yang2016subjective}
R.~Yang, M.~Xu, L.~Jiang, and Z.~Wang, ``Subjective-quality-optimized
  complexity control for {HEVC} decoding,'' in \emph{ICME}, 2016.

\bibitem{yang2016saliency}
R.~Yang, M.~Xu, Z.~Wang, and X.~Tao, ``Saliency-guided complexity control for
  {HEVC} decoding,'' \emph{IEEE TBC}, 2018.

\bibitem{arbelaez2011contour}
P.~Arbelaez, M.~Maire, C.~Fowlkes, and J.~Malik, ``Contour detection and
  hierarchical image segmentation,'' \emph{IEEE TPAMI}, pp. 898--916, 2011.

\bibitem{wallace1992jpeg}
G.~K. Wallace, ``The {JPEG} still picture compression standard,'' \emph{IEEE
  TCE}, pp. xviii--xxxiv, 1992.

\bibitem{He2015Delving}
K.~He, X.~Zhang, S.~Ren, and J.~Sun, ``Delving deep into rectifiers: Surpassing
  human-level performance on imagenet classification,'' in \emph{ICCV}, 2015.

\bibitem{Zeiler2014Visualizing}
M.~D. Zeiler and R.~Fergus, ``Visualizing and understanding convolutional
  networks,'' in \emph{ECCV}, 2014.

\bibitem{kim2016accurate}
J.~Kim, J.~K. Lee, and K.~M. Lee, ``Accurate image super-resolution using very
  deep convolutional networks,'' in \emph{CVPR}, 2016.

\bibitem{bossen2011common}
F.~Bossen \emph{et~al.}, ``Common test conditions and software reference
  configurations,'' \emph{Joint Collaborative Team on Video Coding (JCT-VC),
  JCTVC-L1100}, 2011.

\bibitem{recommendation2002500}
I.~Recommendation, ``500-11,¡°methodology for the subjective assessment of the
  quality of television pictures,¡± {R}ecommendation {ITU-R BT}. 500-11,''
  \emph{ITU Telecom. Standardization Sector of ITU}, vol.~7, 2002.

\bibitem{li2006nonlinear}
D.~Li and X.~Sun, \emph{Nonlinear integer programming}.\hskip 1em plus 0.5em
  minus 0.4em\relax Springer Science \& Business Media, 2006.

\bibitem{zhou201614}
D.~Zhou, S.~Wang, H.~Sun, J.~Zhou, J.~Zhu, Y.~Zhao, J.~Zhou, S.~Zhang,
  S.~Kimura, T.~Yoshimura \emph{et~al.}, ``14.7 {A} 4{G}pixel/s 8/10b {H}.
  265/{HEVC} video decoder chip for 8{K} ultra {HD} applications,'' in
  \emph{ISSCC}.\hskip 1em plus 0.5em minus 0.4em\relax IEEE, 2016, pp.
  266--268.

\bibitem{zhang2016cambricon}
S.~Zhang, Z.~Du, L.~Zhang, H.~Lan, S.~Liu, L.~Li, Q.~Guo, T.~Chen, and Y.~Chen,
  ``Cambricon-x: An accelerator for sparse neural networks,'' in \emph{IEEE/ACM
  MICRO}, 2016.

\bibitem{jouppi2017datacenter}
N.~P. Jouppi, C.~Young, N.~Patil, D.~Patterson, G.~Agrawal, R.~Bajwa, S.~Bates,
  S.~Bhatia, N.~Boden, A.~Borchers \emph{et~al.}, ``In-datacenter performance
  analysis of a tensor processing unit,'' in \emph{ACM ISCA}, 2017.

\end{thebibliography}

\end{document}